\documentclass[fleqn,twoside]{article}
\usepackage[headings]{espcrc2}
\usepackage{amsfonts, amssymb}

\readRCS
$Id: espcrc2.tex,v 1.2 2004/02/24 11:22:11 spepping Exp $
\usepackage{graphicx,color}
\usepackage[figuresright]{rotating}

\newcommand{\AmS}{{\protect\the\textfont2
  A\kern-.1667em\lower.5ex\hbox{M}\kern-.125emS}}
  
\newcommand{\be}{\begin{equation}}
\newcommand{\ee}{\end{equation}}
\newcommand{\bea}{\begin{eqnarray}}
\newcommand{\eea}{\end{eqnarray}}
\newcommand{\f}{\frac}
\newcommand{\rsq}{{\mathfrak R^2}}
\newcommand{\nn}{{\cal N}}
\newcommand{\la}{\label}
\newcommand{\zt}{\tilde{z}}

\newcommand{\cor}[1]{\left\langle{#1}\right\rangle}
\newcommand{\eps}{\varepsilon}

\newcommand{\al}{\alpha}
\newcommand{\bt}{\beta}

\newcommand{\om}{\omega}

\title{Time-dependent AdS/CFT correspondence and the Quark-Gluon plasma}

\author{Alice Bernamonti\address{
Theoretische Natuurkunde, Vrije Universiteit Brussel, and
International Solvay Institutes\\ 
Pleinlaan 2, B-1050 Brussels, Belgium}%
        \thanks{Aspirant FWO; Alice.Bernamonti@vub.ac.be}
        and Robi Peschanski\address{
        Institut de Physique Th\'eorique, CEA, and CNRS (URA 2306), 91191 Gif/Yvette Cedex, France}
        \thanks{robi.peschanski@cea.fr}
        }
       
\runtitle{Time-dependent AdS/CFT correspondence and the Quark-Gluon Plasma}
\runauthor{A. Bernamonti and R. Peschanski}

\begin{document}

\begin{abstract}
Experiments on high-energy heavy-ion collisions reveal the formation and some
intriguing properties of the Quark-Gluon Plasma (QGP), a new phase of matter predicted by Quantum Chromodynamics (QCD), the quantum field theory of strong interactions. The phenomenological success of relativistic hydrodynamic simulations with remarkably weak shear viscosity, modeling QGP as an almost {\it perfect fluid}, are in favor of the occurrence of a strongly-coupled QGP expanding and cooling during the reaction. A derivation of these features in QCD at strong coupling is still lacking and represents a very intricate theoretical challenge. As a quite unique modern tool to relate these dynamical features to a microscopic gauge field theory at strong coupling, time-dependent realizations of the AdS/CFT correspondence provide a fruitful way to study these properties in a realistic kinematic configuration. Relating a 4-dimensional Yang-Mills gauge theory with four supersymmetries (which is a conformal field theory, CFT$_4$) with gravity in Anti-de Sitter space in five dimensions (AdS$_5$), the AdS/CFT correspondence provides a useful ``laboratory'' to study yet unknown strong coupling properties of QCD. Besides the interest of revealing new aspects of the AdS/CFT correspondence in a dynamical set-up, the application to plasma formation leads to non trivial theoretical properties, as we will discuss in the lectures. The highlights of the present lectures are:

1. Emergence of an (almost) perfect hydrodynamic fluid at late proper-times
after the collision.

2. Duality between an expanding 4-dimensional plasma and a black
hole moving radially in the bulk.

3. Intimate link between conformal hydrodynamics and
Einstein's equations in the asymptotically AdS$_5$ space.

4. Possibility of studying the far-from-equilibrium stage of a gauge field
theory at early collisional proper-times.

\vspace{1pc}
\end{abstract}

\maketitle


\section{Heavy-Ion reactions and the QGP}

One of the most striking lessons one may draw \cite {review,hydro} from experiments on heavy-ion collisions at high energy 
($e.g.$ at the RHIC accelerator) is that fluid hydrodynamics seems to be relevant for understanding the dynamics of the reaction 
(see, for instance, the reviews \cite{hydro}).  Indeed, the elliptic flow \cite{JY} describing the anisotropy of the low transverse momentum particles 
produced in a collision at non zero impact parameter implies the existence of a collective 
flow of particles. It agrees with the picture of  an hydrodynamical  pressure gradient due to the 
initial eccentricity in the collision\footnote{
Note that a preliminary analysis of recent LHC data on the elliptic flow at the Alice detector 
seems to confirm  the validity of the hydrodynamic approach \cite{Luzum}.
}. 
Moreover, the hydrodynamic 
simulations which are successful to describe this elliptic flow are consistent 
with an almost ``perfect fluid'' behavior, $i.e.$ a small {\it viscosity over entropy ratio} $\eta/s$ \cite{luzrom}. 
This ratio is particularly interesting since it depends mainly on fundamental features of the fluid, being in particular independent of the particle density (see later, eq.(\ref{visco})).

\begin{figure}[t]
\begin{center}
\includegraphics[width=17pc]{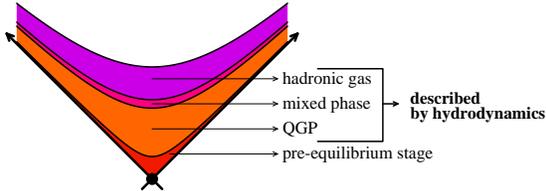}
\end{center}
\caption{\small{\it Description of QGP formation in heavy ion collisions}. The 
kinematic landscape is defined by ${\tau = \sqrt{x_0^2-x_1^2}\ ;\ {\eta=\f 12 
\log \f 
{x_0+x_1}{x_0-x_1}}\ ;\ {x_T\!=\!\{x_2,x_3\}}} $. 
The coordinates along the light-cone are $x_0 \pm x_1,$ the transverse 
ones 
are $\{x_2,x_3\}$ and $\tau$ is the proper time, while $\eta$ is called the ``space-time 
rapidity''.}
\label{1}
\end{figure}

The validity of an hydrodynamical description assuming a quasi-perfect fluid behavior has been nicely anticipated by Bjorken\footnote{The 
introduction of hydrodynamics in the description of  high-energy hadronic collisions has been originally proposed by Landau \cite{lan}, assuming  ``full 
stopping'' initial conditions which result in a non boost-invariant solution or $Landau\ flow$ (see \cite{us} for a unified description of Bjorken and Landau 
flows). However ``full stopping'' initial conditions do not seem to agree with present day data.}  in 
Ref.\cite{Bjorken:1982qr}. The so-called {\it Bjorken flow} is based on the  approximation of an intermediate 
stage of the reaction process consisting in   a boost-invariant quark-gluon plasma (QGP) behaving as a  relativistically expanding fluid. It is  formed  
after a (strikingly rapid) thermalization period and finally decays into hadrons. A schematic description of the process in  light-cone kinematics is shown in   
Fig.\ref{1}. Boost-invariance can be justified  in the central region of the collision since the observed distribution of particles is flat, in 
agreement with the prediction of hydrodynamic boost-invariance, where (space-time) fluid 
and (energy-momentum) particle rapidities are proven to be equal \cite{Bjorken:1982qr} (see later the discussion in section 3).

The Bjorken flow was instrumental for deriving and predicting many qualitative and even semi-quantitative 
features of   the quark-gluon plasma formation in heavy-ion reactions. It was later confirmed and developed 
through numerical simulations where various effects have been included, up to the full 4-dimensional structure of the hydrodynamic flow. 
However, since hydrodynamics is an effective mean-field  approach, it says only 
little on the relation with the microscopic gauge field theory, $i.e.$, in the present case, Quantum 
Chromodynamics (QCD). Some important questions  remain unsolved, such as  the reason why the fluid 
behaves like a (almost) perfect fluid, what is the small amount of viscosity it may require,  why and how fast thermalization proceeds, etc...

The problem is made even more difficult by the  strong coupling regime of QCD, which is very probably 
required, since a perturbative description leads in general to  high  $\eta/s$ ratio. 
In fact, large collision cross-sections with transversally cut-off momenta are required in order to explain a small macroscopic 
shear viscosity. The mean free path and the average transverse momentum  induced by the gauge field medium 
should be small, in order to damp  the nearby force  transversal to the 
flow measuring shear viscosity. In a classical physics context, one may write
\be
\frac \eta s=\f{\rho_{part} \times \lambda_{free} 
\times \bar p_{av}}{\rho_{part} \times k_B}\ \left[\ge  {\frac 1{4\pi} \frac \hbar{k_B}}\right]\ ,
\label{visco}
\ee 
where $\rho_{part},$ ($resp.$ $\bar p_{av}$) is the particle density, ($resp.$ average transverse momentum) and $\lambda_{free}$ the mean free-path. 
Both small mean free-path and low average transverse momentum are expected in the non-perturbative  regime of QCD and more generally from a gauge theory at strong coupling. We also quote in formula (\ref{visco}) the AdS/CFT value \cite{son} $1/{4\pi}$ in terms of the fundamental constants' ratio $\hbar/{k_B},$ corresponding to the large coupling limit of the conformal field theory. This  AdS/CFT ratio appears to be  very small compared to all known non-relativistic fluids, as suggested by the inequality  between square brackets.  Indeed, the AdS/CFT value has been proposed as an absolute  theoretical lower bound \cite{son1}. A discussion is in progress to fully confirm this statement or to find counter-examples in microscopic theories (see \cite{universalbound} and references therein). 

In any case, a very small phenomenological viscosity  ratio is confirmed; not meaning that viscosity effects should be neglected . Observations on the elliptic flow primarily, and other observables, allow one to give estimates of $\eta/s$ which are within range of the AdS/CFT value (see, $e.g.$ \cite{Luzum,luzrom}).

It is thus interesting to use our modern (while still in progress) knowledge
of non-perturbative methods in quantum field theory to fill the gap between the 
macroscopic and microscopic descriptions of the quark-gluon plasma produced in 
heavy-ion collisions. Lattice gauge theory methods are very useful to analyze 
the static properties of the quark-gluon plasma, but they are still unable 
to describe the plasma in collision. 

Hence we are led to rely upon the new tools offered by the Gauge/Gravity correspondence and in particular on
 the most studied and well-known of its realizations, namely the AdS/CFT duality \cite{adscft} between 
${\nn}\!=\!4$ supersymmetric Yang-Mills theory and type IIB superstring in the 
large $N_c$ approximation. The properties of the gauge theory in (physical) 
Minkowski space in $3\!+\! 1$ dimensions are in one-to-one 
relation with properties of the bulk theory, in the 10D target space of 
strings. For our applications, strong coupling features will be related by duality 
to the geometric structure of the 5D metric with Minkowski boundary, which is built on top of vacuum AdS. 
At weak coupling, the physical space corresponds to the world-volume of a stack of infinitely many ($N_c\!\to\!\infty$) joint D3-branes.
Note that the overall consistency of the correspondence scheme will be required, even if only geometrical properties will be used in practice.

One should be aware, when using AdS/CFT tools, that there does not yet exist a 
gravity dual construction for QCD. Also, the relation between perturbative (weak-coupling) 
and non-perturbative (strong-coupling) aspects of QCD are left over in the correspondence, which conveniently 
describes the strong coupling regime of gauge theories uniquely. 
However, the nice feature of the quark-gluon plasma is that it is 
a deconfined phase of QCD, characterized by collective degrees of freedom. 
Thus one may expect to get useful information from AdS/CFT. This has 
already been proved in the description of static geometries, by the 
evaluation of $\eta/s$  \cite{son}. Other aspects of the QGP can be studied 
via the duality, in a static plasma configuration at fixed temperature. 

Our specific aim and  the subject of the present lectures is the investigation of the 
Gauge/Gravity duality in a dynamical {\it time-dependent} setting using the AdS/CFT correspondence.  
The goal is to describe a collision process in a strongly coupled gauge theory, 
including in particular the cooling of temperature accompanying the  proper-time evolution of the fluid. 

In section \ref{AdS/CFTstrong}, we introduce the theoretical tool of holographic renormalization and apply it, 
as a warm-up exercise, to a static uniform plasma. Sections \ref{LateTime} and \ref{EarlyTime} present, respectively, 
the late and early proper-time analysis of the plasma evolution, with particular emphasis on boost-invariant flows. 


\section{AdS/CFT and strong gauge interactions}\label{AdS/CFTstrong}

In the previous section, we mentioned the ubiquity of hydrodynamic methods in 
the description of QGP produced at RHIC. Yet, despite their success in 
describing the data, we have to keep in mind that they are used as a 
phenomenological model, without a real derivation from gauge theory. This is 
understandable, since almost perfect fluid hydrodynamics is intrinsically a 
strong coupling phenomenon, for which one lacks a purely gauge theoretical 
method.\footnote{
Lattice QCD methods do not work well here, as this would require 
analytical continuation to Minkowski signature which is nontrivial in this 
context \cite{Eucli}.
} 

On the other hand, there exists a wide class of gauge theories, which can be 
studied analytically at strong coupling. These are superconformal field theories 
with gravity duals. In the unifying context of string theory, using the AdS/CFT correspondence, one is able to map 
gauge theory dynamics (CFT) at strong coupling and large number of colors into 
solving Einstein's equations in asymptotically anti-de Sitter space (AdS). The 
theories with gravity duals can differ substantially from real world QCD at zero 
temperature. The best known example of such theories - $\mathcal{N} = 4$ super 
Yang-Mills (SYM) - is a superconformal field theory with matter in the adjoint 
representation of the gauge group SU(N$_{c}$). Because of the conformal symmetry 
at the quantum level this theory does not exhibit confinement. On the other hand, 
differences between $\mathcal{N} = 4$ SYM and QCD are less significant above 
QCD's critical temperature, when quarks and gluons are in the deconfined phase. 
Moreover, it was observed on the lattice that QCD exhibits a quasi-conformal 
trend \cite{Lattice} in a range of temperatures $T > 300$ MeV. There the equation of state is 
reasonably approximated by the conformal relation $\epsilon = 3 p$, corresponding to a traceless energy-momentum tensor.
The above observations, together with experimental results suggesting that QGP is a strongly coupled 
medium, are an incentive to use the AdS/CFT correspondence as a tool to get insight into non-perturbative dynamics.


\subsection{AdS/CFT correspondence}

We will now describe how to set up an AdS/CFT computation for determining the 
space-time behavior of the gauge field plasma corresponding to a Bjorken boost-invariant flow. 
The  problem is formulated in terms  of the proper-time evolution of the matter energy-momentum tensor on the CFT boundary \cite{Janik:2005zt}. 
This method does not make any underlying assumption about local equilibrium or 
hydrodynamical behavior. We will indeed obtain the hydrodynamic expansion as a generic consequence of the late 
time behavior of the expanding and strongly coupled plasma. 

Suppose we consider some macroscopic state of the plasma characterized by a 
space-time profile of the energy-momentum tensor
\be
T_{\mu\nu}(x^\rho)\ ,\quad \mu, \nu, \rho =\{0,\cdots,3\}\ .
\ee
Then, since the AdS/CFT correspondence asserts an exact equivalence between gauge 
and string theories, such a state should have its counterpart on the gravity side 
of the correspondence. Typically, it will be given by a modification of the 
geometry of the original $AdS_5\times S^5$ metric. This follows from the fact that 
operators in the gauge theory correspond to  fields in supergravity (or string 
theory). When we consider a state with a nonzero expectation value of an 
operator, the dual gravity background will have the corresponding field 
modified from its `vacuum' value. In the case of the energy-momentum tensor, 
the corresponding field is just the 5D metric. One then has to 
assume that the geometry is well defined,  $i.e.$ it is free from naked 
singularities - singularities not hidden by an event horizon -.
This principle selects the allowed physical space-time profiles of the 
gauge theory energy-momentum tensor. Thus 
this becomes the main dynamical mechanism for selecting the appropriate strongly coupled gauge theory from the dual Einstein equations.
In practice, the geometry will be obtained from  a ``holographic renormalization'' procedure starting from the boundary. 
The boundary conditions serve as initial conditions for the  construction of the bulk features, the fifth dimension acting as a renormalization scale. 
Let us describe in detail this construction.


\subsection{Holographic renormalization}\label{HR}

The Gauge/Gravity  duality can be described as an ``holographic'' correspondence between the 4-dimensional physical 
space where the gauge theory lives and the 5-dimensional space where the supergravity (weak curvature) approximation 
of the 10-dimensional string theory is valid. It means qualitatively that the whole information should be the same on both 
sides of the correspondence, despite the difference in dimensionality. 
In practice, this notion of ``holography'' has a precise realization in terms of the ``holographic renormalization''
program \cite{Skenderis}. Let us illustrate the holographic renormalization by simple examples, which will in fact be sufficient for the applications we have in mind.

Suppose that, when considering the presence of matter on the 4-dimensional physical space, our corresponding 5-dimensional geometry is parameterized by
\be
ds^2=\f{g_{\mu\nu}(x^\rho,z) dx^\mu dx^\nu +dz^2}{z^2} \equiv g^{5D}_{\al\bt} dx^\al dx^\bt,
\la{ffg}\ee
where we adopt the Fefferman-Graham definition \cite{fg} of the 5D metric. 
The flat case $g_{\mu\nu} = \eta_{\mu\nu}$ parametrizes AdS$_5$ in Poincar\'e coordinates. The conformal boundary of space-time is at $z\!=\!0$.

Considering the general metric (\ref{ffg}) from the point-of-view of the AdS/CFT correspondence, the  following questions are in order:\newline

i) What are the constraints imposed on $g_{\mu\nu}(x^\rho,z)$?\\

ii) What is the corresponding energy-momentum profile $\cor{T_{\mu\nu}(x^\rho)}$?\\


The metric (\ref{ffg}) has to be a solution of 5D Einstein's equation with negative 
cosmological constant\footnote{
One can show that such solutions lift to 10D 
solutions of ten dimensional type IIB supergravity. The effective 5D negative 
cosmological constant comes from the 5-form field in 10D supergravity.
}
\be
\label{e.einst}
R_{\alpha\beta}= \f{1}{2}g_{\alpha\beta} \left( R - 12\right) \ .
\ee
The expectation value of the energy momentum tensor may be recovered by 
expanding the metric near the boundary $z\!=\!0$. The ``holographic 
renormalization'' procedure \cite{Skenderis}, for a Minkowskian boundary metric, gives\footnote{
In principle, in (\ref{bound1}) also additional logarithmic terms appear (see \cite{Skenderis}). They are absent in all the cases we are going to examine.}
\be
g_{\mu\nu}(x^\rho,z)=\eta_{\mu\nu}+z^4 g^{(4)}_{\mu\nu}(x^\rho)+\ldots\ ,
\la{bound1}
\ee
and
\be
\cor{T_{\mu\nu}(x^\rho)} = \f{N_c^2}{2\pi^2} \cdot g^{(4)}_{\mu\nu}(x^\rho)\ .
\la{bound2}
\ee
The relations   (\ref{bound1},\ref{bound2}) can be used in two ways. Firstly, given a solution of Einstein's 
equations, we may read off the corresponding gauge theoretical energy-momentum 
tensor. Secondly, given a traceless and conserved energy-momentum profile, one 
may integrate Einstein's equations into the bulk in order to obtain the dual 
geometry\footnote{
This can be done order by order in $z^{2}$, which is a 
near-boundary expansion. However potential singularities are hidden deep in the 
bulk, thus this power series needs to be resummed.
}. 
Then the criterion of non singularity of the geometry will determine the allowed 
space-time evolution of the plasma. Let us note that this formulation is in fact 
quite far from a conventional initial value problem.


\subsection{Example: static uniform plasma}

Before moving to the case of an  expanding plasma, it is convenient to consider the 
simple situation of a static uniform plasma with a constant energy-momentum 
tensor. This will provide a useful exercise introducing the main tools for further applications.\\ \\

Starting from a constant diagonal energy-momentum tensor
\be
T_{\mu\nu}\! = \!
\left(\begin{tabular}{cccc}
$\epsilon$ & 0 &0 & 0 \\
0 & $p_\parallel$ & 0 & 0 \\
0 & 0 & $p_\perp$ & 0 \\
0 & 0 & 0 & $p_\perp$
\end{tabular}\right)\,,
\ee 
with $p_\parallel=p_\perp$ and energy density and pressure related by $\epsilon=3p$,
one has to solve the 5D Einstein's equations for an Ansatz
\be
ds^2=\f{-e^{a(z)} dt^2
+e^{b(z)} {d\vec x^2}+ {dz^2}}{z^2}
\la{An}\ee
obeying the above-mentioned boundary conditions (\ref{bound1},\ref{bound2}). 
Inserting (\ref{An}) into the Einstein's equations (\ref{e.einst}) and splitting them into components, one gets\\ \\
$R_{11}\Rightarrow 
-8a'+za'^2-6b'+3za'b'+2za''=0\ ,$\\
$R_{55}\Rightarrow -2a'+za'^2-6b'+3zb'^2+2z(a''+3b'')=0\ ,$
$R_{22}\Rightarrow 
-2a'-12b'+za'b'+3zb'^2+2zb''=0\ .$\\ \\
Then
\be
R_{11}-R_{55}\Rightarrow a'=\f{zb'^2+2zb''}{zb'-2}\,, 
\ee
while combining it with the expression from $R_{22}$ for $a'$ gives
\be
-3b'^2+zb'^2+zb''=0\,. 
\ee
From those last two equations, it is easy to obtain independent equations for $a(z)$ and $b(z)$ and their solutions.  With appropriate normalizations at $z\!=\!0,$ one gets
\bea 
b&\equiv& \log{\left(1+\f{z^4}{z_0^4}\right)} \nonumber\\
a&\equiv& 2\log{\left(1-\f{z^4}{z_0^4}\right)}-\log{\left(1+\f{z^4}{z_0^4}\right)}\ ,
\eea
where $z_0$ is the integration constant. Hence, Einstein's equations can be  exactly solved analytically in this case and we find 
\cite{Janik:2005zt}
that the exact dual geometry of such a system is
\bea
\label{e.bhfef}
ds^2=-\f{\left(1\!-\!\f{z^4}{z_0^4}\right)^2}{\left(1\!+\!\f{z^4}{z_0^4}\right)} \f{dt^2}{z^2}\! +\! \left(1\!+\!\f{z^4}{z_0^4}\right) \f{d\vec x^2}{z^2}\! + \! \f{dz^2}{z^2}.
\eea 
This metric may look at first glance unfamiliar, but a change of coordinates
\be
\zt \equiv \f{z}{\sqrt{1+\f{z^4}{z_0^4}}}
\ee
transforms it to the standard metric form of the AdS-Schwarzschild static black hole\footnote{
The identification of the metric (\ref{e.bhfef}) as the dual of a static fluid at fixed temperature has previously  been made in \cite{bala} using the  approach `from bulk to boundary'. 
This is the reverse way w.r.t. holographic renormalization which proceeds `from  boundary to bulk'. The interest of holographic renormalization is that it allows a construction 
of the dual geometry starting  only from 4-dimensional physical data.
}
\bea
\zt^2 ds^2=-\left(1-\f{\zt^4}{\zt_0^4}\right) dt^2\! + d\vec x^2\! + \f{1}{1-\f{\zt^4}{\zt_0^4}} d\zt^2,
\label{standard}\eea
with $\zt_0=z_0/\sqrt{2}$ being the location of the horizon. 
Before we proceed further, let us note here one crucial thing: the fact that the dual geometry of 
a gauge theory system with constant energy density is a black hole was {\em not} 
an assumption, but rather an outcome of a computation.

The Hawking temperature
\be
T=\f{1}{\pi \zt_0} \equiv \f{\sqrt{2}}{\pi z_0}
\la{haw}\ee
is then identified with the gauge theory temperature, and the entropy with the 
Bekenstein-Hawking black hole entropy 
\be
S=\f{N_c^2}{2\pi\zt_0^3}=\f{\pi^2}{2} N_c^2 T^3
\ee
which is $3/4$ of the entropy at zero coupling. This identification between the black hole characteristics and 
the equilibrium thermodynamics of the plasma reveals a striking correspondence between the near-horizon 
and boundary properties of the gauge/gravity system.

To finish our discussion of the static black hole, we note that the Fefferman-Graham coordinates cover only the 
part of space-time lying outside of the horizon. It will appear useful to introduce different coordinates allowing 
one to go inside the horizon, such as the Eddington-Finkelstein's coordinate system. 
Through the time coordinate redefinition
\be
u \equiv t-\f{1}{4} \tilde{z}_0 \left( 2 \arctan \f{\zt}{\tilde{z}_0} +
{\log \f{\tilde{z}_0+\zt}{\tilde{z}_0-\zt}} \right)\ ,
\ee 
the metric (\ref{standard}) becomes:
\be
\zt^2 ds^2=- \left(1-\f{\zt^4}{\tilde{z}_0^4} \right) du^2  +d\vec x^2 + 2 du d\zt \ . 
\ee 
Now the metric is well defined at $\zt=\tilde{z}_0$ and the horizon can be smoothly crossed. 
Note that the time coordinate gets an infinite shift near the horizon, here being harmless, but which reveals to be 
quite subtle for the time-dependent backgrounds we will consider now.


\section{AdS/CFT and late time quark-gluon plasma flow}\label{LateTime}

In this part, we describe applications of the methods introduced in section \ref{AdS/CFTstrong} 
to the analysis of plasma dynamics in strongly coupled gauge theories. The time dependence of the system introduces considerable 
computational difficulties and it is in general a hard task to achieve a detailed quantitative description of the plasma evolution. 
Nevertheless, significant information on the dynamics can be extracted working in simplified and 
symmetric, but realistic setups. 
We shall  consider specific regimes of the problem in exam, but some general lessons can still be drawn about the highly nonlinear regime of the plasma expansion. 

Throughout the next sections we will model the products of heavy-ion collisions with a $\mathcal{N}\!=\!4$ SYM plasma at finite 
temperature. We mainly focus on the results obtained  in Ref.~\cite{Janik:2005zt} concerning the late proper-time behavior of the quark-gluon plasma. Recent and detailed reviews on the subject are also  Refs.~\cite{Heller:2008fg} and  \cite{Janik:2010we} . 

The structure of the sections is organized in the following way. In section \ref{Bjorkenflow} we review briefly the ``Bjorken flow" description of the central rapidity region 
of heavy-ion collisions. This serves as an introduction to section \ref{Boost-invFlow}, where we present the gauge theory setup which has been used in 
\cite{Janik:2005zt} to model the products of heavy-ion reactions. The description is in fact inspired by the hydrodynamic description of heavy-ion collisions by  Bjorken \cite{Bjorken:1982qr} and based on the central assumption of boost invariance. This hypothesis seems experimentally to be valid in the central rapidity region of the heavy-ion collision process.
In section \ref{LargeTime}, we review the computation of the holographic dual geometry of the plasma system in the asymptotic limit of large proper-time.
The properties of this class of time-dependent space-times are investigated in section \ref{PerfectFluid}. 
The requirement of the absence of naked singularities in the bulk selects uniquely the gravity dual of a perfect 
fluid. This non-singular solution can be interpreted as a black hole moving off from the AdS boundary in the radial direction, as a function of proper-time. We conclude 
with sections \ref{Viscosity&Relaxation} and \ref{BeyondBoostInv}, where we present attempts of going beyond the assumptions of perfect hydrodynamics and boost invariance.


\subsection{Hydrodynamic Bjorken flow}\label{Bjorkenflow}

The introduction of relativistic hydrodynamics in the description of high-energy hadronic reactions is due to Landau in a remarkable premonitory paper \cite{lan}. However, the experimental evidence of a connection between heavy-ion collisions and relativistic hydrodynamics has been first modeled 
by Bjorken in \cite{Bjorken:1982qr}. In this description, boost-invariance of the plasma arises as an outcome of the assumption
of hydrodynamics, in a semi-classical approximation where the fluid follows ballistic trajectories or, in other terms, the momentum-space rapidity is equal to the space-time one (see Fig.\ref{Inout}). 
The model is therefore well suited to parametrize the central region of ultra-relativistic 
collisions, for which the particle distribution is nearly flat as a function of rapidity. 
\begin{figure}[t]
\centerline{\includegraphics[width=.4\textwidth]{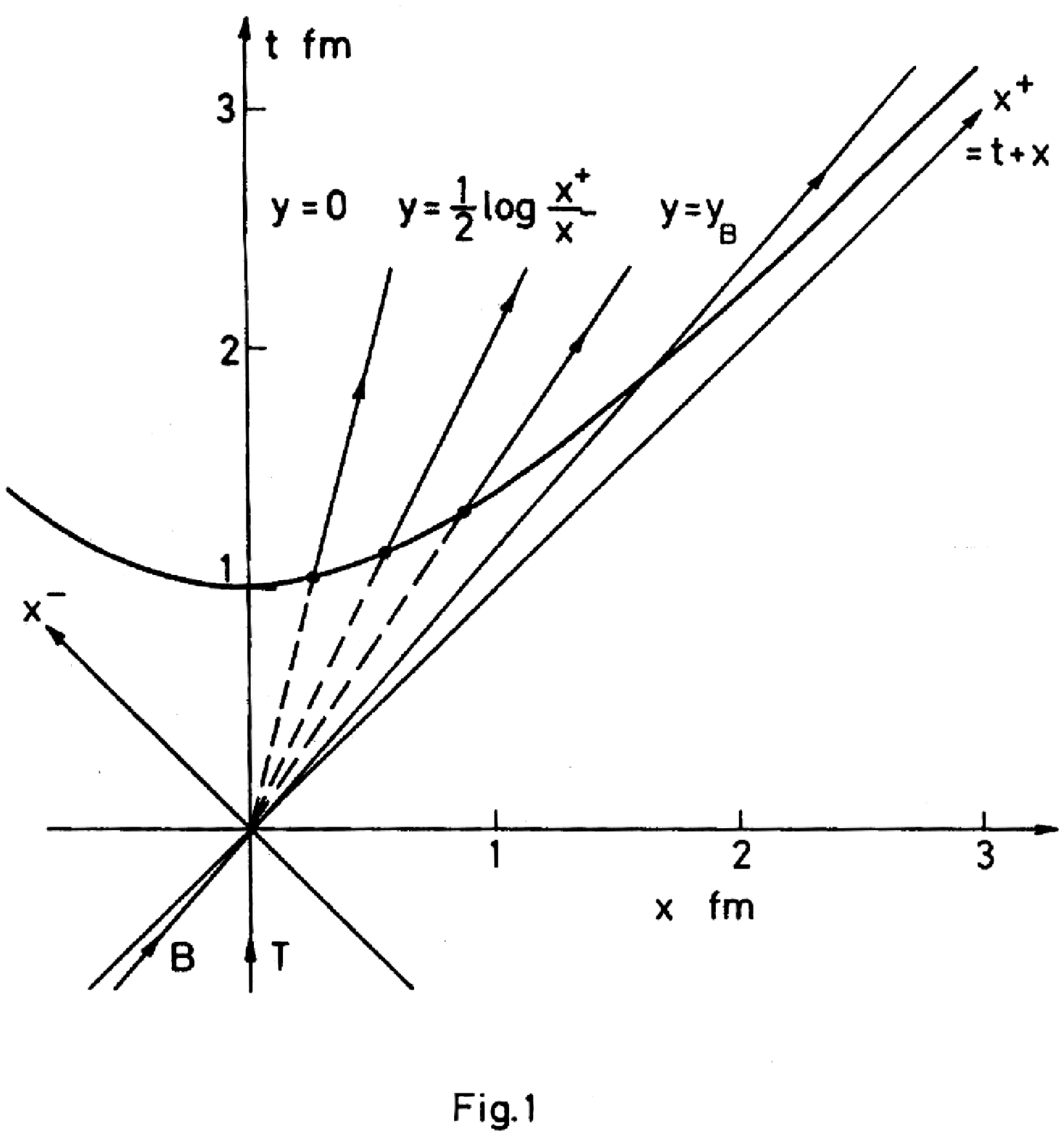}}
\caption{Semi-classical fluid trajectories: the In-Out cascade. The rapidity $y$ is equal to the space-time rapidity $\eta =\f 12 \log \f {x^+}{x^-}.$}
\la{Inout}
\end{figure}

In this section, we review a number of aspects of this picture. The summary is intended to be useful in the follow-up to motivate the assumptions of the 
AdS/CFT models we discuss and to interpret the physical results we obtain.

For the discussion that follows, it is convenient to parametrize flat four-dimensional space-time of coordinates $(t,x_1,x_{\perp} )$, where 
$x_{\perp}= (x_2, x_3)$, in terms of the light-cone coordinates $(x^+ , x^-, x_{\perp})$. The latter are defined by $x^{\pm} \equiv t \pm x_1$, in which 
the flat space-time metric reads 
\be \label{lightcone}
ds^2 = - dx^+ dx^- + d x_{\perp}^2\,.
\ee
The proper-time $\tau$ and space-time rapidity $\eta$ of the fluid are related to the light-cone coordinates by $x^{\pm} \equiv \tau \, e^{\, \pm \eta}$.

In the first stages of the collision, a rather dense interacting medium is created and the individual partonic or hadronic degrees of freedom can be 
neglected to a good approximation. This observation justifies the treatment of the medium as a fluid and, moreover, allows the assumption of local equilibrium. In the simplest perfect fluid 
approximation, the energy-momentum tensor reads
\be \label{perfectfluid}
T^{\mu \nu}= (\epsilon + p) u^{\mu} u^{\nu} - p \, \eta^{\mu \nu}\,,
\ee
where $\epsilon$,  $p$ and $u^{\mu}$ are the energy density, pressure and 4-velocity of the fluid, respectively; while $\eta^{\mu \nu}$ is the flat metric of expression (\ref{lightcone}). 
We further assume an equation of state of the form
\be\label{eqstate}
\epsilon = g\, p\,,
\ee
where $1/\sqrt{g}$ is the sound velocity of the liquid. In the case of a conformal theory, the traceless condition $T_{\mu}^{\mu}=0$ implies $g =3$ in four dimensions. 
The continuity equation $\partial_{\mu} T^{\mu \nu}=0$ for a perfect fluid of energy density and pressure related by (\ref{eqstate}) reduces to two equations\\
\bea \label{continuityeq}
g \partial_+(\ln p)&=&\!\!\!\!\! -\frac{(1+ g)^2}{2}\partial_+ p - \frac{g^2\!-\!1}{2} e^{-2 y} \partial_- y \nonumber \\
g \partial_-(\ln p)&=& \frac{(1+ g)^2}{2}\partial_- p - \frac{g^2\!-\!1}{2} e^{2 y} \partial_+ y  
\eea
for the pressure and the rapidity 
\be
y \equiv \frac{1}{2}\ln \left(\frac{\epsilon+p}{\epsilon -p} \right)
\ee
of the fluid. The other thermodynamical quantities can be derived from the equation of state (\ref{eqstate}) and the standard thermodynamical identities
\be
p+ \epsilon = T s \,, \qquad d\epsilon = T d s\,,
\ee
where $T$ and $s$ are the temperature and entropy density of the fluid and we have assumed, for simplicity, vanishing chemical potential. 

The result is
\be
 p= \f{\epsilon }{g} = p_0 \, T^{g+1}\,, \quad s = s_0 \, T^{g} \propto p^{\f{g}{g+1}}\,,
\ee
for constant $p_0$ and $s_0$. 

The key observation is that the expansion of the fluid can be described by a semi-classical picture.  In fact, the high occupation numbers of the system allow to assume 
that the fluid components follow quasi-classical trajectories in space-time, here straight-line trajectories starting at the origin (see Fig.\ref{Inout}), and given by
\be \label{Bjorkenansatz}
y= \eta\,.
\ee
Plugging the Ansatz (\ref{Bjorkenansatz}) into (\ref{continuityeq}), one obtains
\be
g \partial_+(\ln p)= -\frac{1+ g}{2 x^+}\,, \quad g \partial_-(\ln p)=- \frac{1+ g}{2 x^-}\,,
\ee
from which it easily follows
\be
\ln \f p{p_0} = \f{1\!+\!g}2\log x^+x^-\ \Rightarrow \ p(\tau) = \f{p_0}{ \tau^{\f{1+g}{g}}}\ .
\ee
The simple Ansatz  (\ref{Bjorkenansatz}) thus leads to a boost-invariant system. Such a physical description is suitable to 
model the central rapidity region of highly relativistic heavy-ion collisions, where a central plateau in the distribution of particles is detected.

Specifically, in the perfect fluid four-dimensional case, the behavior is
\be \label{BjorkenFlow}
p(\tau) = \frac{\epsilon(\tau) }{3}  =\frac{p_0}{ \tau^{\f{4}{3}}} \sim T^{\,4}(\tau)\,, \quad s\sim T^3(\tau)\ .
\ee
Note that the entropy density verifies $s =s_0 \tau^{-1}$. Hence the total entropy $S$ per unit of rapidity (and transverse area) is constant \cite{Bjorken:1982qr} (as expected from a perfect fluid) since
\be
dS \equiv \int s \, d^3x = s \tau \int d^2x_{\perp} dy\ . 
\la{entropy}
\ee 


\subsection{Boost-invariant flow}\label{Boost-invFlow}

Inspired by  Bjorken's analysis,  the methods provided by the AdS/CFT correspondence have been applied to study 
the dynamics of boost-invariant strongly interacting gauge theory matter in \cite{Janik:2005zt}. There are technical reasons for making this symmetry assumption, due to the complication of the Einstein's equations in the gravity dual. On a more theoretical level, using the $\mathcal N =4$ SYM theory as a substitute to QCD seems at first sight highly problematic. Moreover,  experimentally  the assumption 
of boost invariance is not optimal, since the observed multiplicity distribution of particles in heavy-ion collisions is rapidity-flat only in the central region. In terms of the hydrodynamic modelization, this means that the entropy is rapidity-independent, and thus obeying boost-invariance, only in the same region.

All these warning remarks have to be certainly taken into account, when using the AdS/CFT correspondence in a physical context. However, there are convincing arguments to show that for appropriate observables and by using AdS/CFT methods, one obtains a first fruitful approach to the behavior of QCD at strong coupling. From a phenomenological point-of-view, boost-invariance  is  widely used in modeling 
the relativistic heavy-ion collisions at RHIC in the central region of rapidity, where the hydrodynamic approach seems successful in describing the data and the occurrence of the quark-gluon plasma. Moreover, this plasma is a non-confining phase of QCD for which no specific hadronic scale seems to be relevant. In a more theoretical language, in this regime QCD appears to be  nearly conformal. The above arguments suggest that it is reasonable to apply boost-invariance, at least in a first approximation, and to study the dynamics 
of the $\mathcal N =4$ SYM plasma under consideration.

In a boost-invariant setup, it is natural to parametrize the flat four-dimensional boundary in terms of the coordinates $(\tau, \eta, x_{\perp})$, 
where $\tau$ is the proper-time, $\eta$ the space-time rapidity and $x_{\perp} = (x_2,  x_3)$ denotes the transverse directions. We already introduced 
these coordinates in the previous section, but it is worth being more explicit here. In terms 
of the Minkowski coordinates $(t,x_1,x_{\perp})$, the definition of $\tau$ and $\eta$ reads
\be
t = \tau \cosh \eta \,, \qquad x_1 = \tau \sinh \eta \,,
\ee
and the metric has the form
\be
ds^2 = - d\tau^2 + \tau^2 d\eta^2 + dx_{\perp}^2\,.
\ee
For simplicity, further assume independence on the transverse coordinates $x_{\perp}$, corresponding to the limit of infinitely large nuclei colliding. 

The energy-momentum tensor $T_{\mu \nu}$ of the gauge theory can then be written in diagonal form and has only three non-vanishing components: $T_{\tau \tau}$, 
$T_{\eta \eta}$ and $T_{x_2 x_2} = T_{x_3 x_3}$, which depend on proper-time. The conservation condition $D_{\nu}T^{\mu \nu} =0$ 
and, since we are dealing with a conformal theory, the tracelessness condition $T_{\mu}^{ \mu}=0$ further restrict the form of the energy-momentum tensor.  
It can be shown that all components can be expressed in terms of a single function $\epsilon= \epsilon (\tau)$:
\bea \label{Tboost}
T_{\mu \nu} =\textrm{diag}\left ( \epsilon,  -\tau^2(\tau \epsilon^{\prime}\! + \!\epsilon) ,   \epsilon \!+  \!\f{ \tau \epsilon^{\prime}}{2} , \epsilon\! +\!  \f{\tau \epsilon^{\prime}}{2}  \right ).
\eea
The function $\epsilon(\tau)$ can be interpreted as the energy density of the plasma at mid-rapidity ($i.e.$ at $x_1 =0$) as a function of proper-time. 
The last kinematic constraint on the stress-energy tensor is the positive energy condition. For any time-like vector $t^{\mu}$, the energy density in the reference frame 
whose time-like direction is specified by $t^{\mu}$ should be non-negative
\be\label{positiveenergy}
T_{\mu \nu}t^{\mu} t^{\nu} \ge 0\,.
\ee
For the Ansatz (\ref{Tboost}), it implies
\be \label{epsiloncond}
\epsilon(\tau) \ge 0\,, \quad \epsilon^{\prime}(\tau) \le 0\,, \quad \tau \epsilon^{\prime}(\tau)\ge - 4 \, \epsilon(\tau)\,.
\ee
The explicit expression of $\epsilon$ specifies the dynamics of the theory and it is the information we aim to determine through the AdS/CFT correspondence. 

In the following, we restrict the analysis to the class of gauge theories whose late proper-time evolution is specified by a power-law parameter $s,$ such that
\be \label{fboost}
\epsilon (\tau) \propto \f{1}{\tau^{s}}\,.
\ee
The positive energy condition (\ref{positiveenergy}) fixes the range of possible powers $s$ to $0 \le s \le 4$.

Notice that the family of gauge theories described by (\ref{fboost}), includes a number of cases of physical interest. If the longitudinal pressure $T_{\eta \eta}$ vanishes, see (\ref{Tboost}), the plasma 
is in a {\it free streaming} phase which is expected to be well suited to the case of a weakly coupled plasma \cite{Kovchegov:2005ss}. Solving for $\epsilon$ in (\ref{Tboost}) leads to
$\epsilon(\tau) \propto \tau^{-1}$. If the gauge theory plasma behaves, instead, as a perfect fluid, the energy-momentum tensor has to be of the form (\ref{perfectfluid}) with $\epsilon= 3\, p$. 
Under the assumption of boost invariance, $u^{\mu}= (1,0,0,0)$ and the comparison with (\ref{Tboost}) leads to the Bjorken solution of section \ref{Bjorkenflow}:  $\epsilon(\tau) \propto \tau^{-4/3}$. 
The case $\epsilon(\tau) \propto \tau^{0}$ is also of interest, it describes a fully anisotropic medium with negative longitudinal pressure $p_\parallel \equiv T_{\eta \eta}/\tau^2=-\epsilon=-p_\perp$ that might be relevant for the early times dynamics at strong coupling, as advocated in \cite{Kovchegov:2007pq}. As we shall see in section \ref{EarlyTime}, the status of  early times dynamics  in the AdS/CFT correspondence seems to be more complex.

While we will deal in detail with the dual description of the early-time flow in the following, in the present part we restrict the analysis to the range $0< s< 4$.
In what follows, we will determine which, if any, of the above behaviors is relevant for the description of the large proper-time evolution.


\subsection{Large proper-time behavior of the gravity dual}\label{LargeTime}

In this section, following the procedure of holographic renormalization introduced in section \ref{HR}, we repeat the construction of \cite{Janik:2005zt} of the dual geometries to the boost invariant gauge 
theories with energy densities of the form (\ref{fboost}). Since the bulk metric shares the same symmetries of the plasma system, the most general Ansatz in Fefferman-Graham 
coordinates reads
\bea\label{FGansatz}
z^2 ds^2 &= &-e^{a(\tau,z)} d\tau^2 +\tau^2 e^{b(\tau,z)} d\eta^2\nonumber\\
&&  + \, e^{c(\tau,z)} dx^2_{\perp} +dz^2\,.
\eea
According to the holographic dictionary,  to determine the metric components $a(\tau,z)$, $b(\tau,z)$ and $c(\tau,z)$, 
Recalling Einstein equations with a negative cosmological constant
\be\label{Einstein'seqs}
R_{\alpha \beta} = \frac{1}{2}g_{\alpha \beta} (R -12) 
\ee
one has to solve them in a power series expansion in the radial coordinate $z$.
The indices $\alpha, \beta$ in (\ref{Einstein'seqs}) denote the bulk coordinates $(\tau,z, \eta, x_{\perp})$ and the metric components are subject to the boundary conditions 
\be
a(\tau,z) = - z^4 \epsilon(\tau) + z^6 a_6(\tau) + z^8 a_8(\tau) + \dots
\la{BCa}\ee
around $z\!=\!0$ and similar ones for $b(\tau,z)$ and $c(\tau,z)$.
By studying the explicit form of the Fefferman-Graham expansion, it can be shown that the large 
$\tau$ asymptotics of the exact solutions can be obtained analytically (see \cite{Janik:2005zt} for the details of the computation). 
In fact, introducing the scaling variable
\be \label{scalingvariable}
v = \f{z}{\tau^{\f{s}{4}}}\,,
\ee
the exact solutions have an expansion of the form
\be
a(\tau,z) = a(v) + {\mathcal O} \left(\frac{1}{\tau^{\alpha}} \right)\,,
\ee
for some positive power $\alpha$. 
The asymptotic equations, obtained from  (\ref{Einstein'seqs}) taking the limit $\tau \to \infty$, while keeping $v$ fixed, can then be solved exactly.
The solution is 
\bea
a(v) &=& A(v)-2m(v) \nonumber \\
b(v) &=& A(v)+(2s-2) m(v) \\
c(v) &=& A(v)+(2-s) m(v) \nonumber 
\eea
where
\bea
A(v) & = & \f{1}{2} \ln \left(1- \Delta^2(s) v^8\right) \label{A&m}  \nonumber \\ 
m(v) & = & \f{1}{4\Delta(s)}\ln\frac{1+\Delta(s) v^4}{1-\Delta(s) v^4}\\
\Delta(s) & = &   \sqrt{\f{3\, s^2-8s+8}{24}}\,. \nonumber 
\eea


\subsection{Selection of the perfect fluid}\label{PerfectFluid}

Let us now examine in more detail the geometry we obtained.
The dual asymptotic metric of the boost-invariant plasma has a potential singularity for $v = \Delta^{-1/4}(s)$, where the argument of the logarithm in (\ref{A&m}) 
vanishes. 

As usual, to distinguish a physical singularity from a coordinate one, it is sufficient to construct a scalar invariant formed out of the Riemann tensor which diverges at 
the singularity location. While the curvature scalar is everywhere regular, the simplest non-trivial scalar that can be considered is the square of the Riemann tensor (or Kretschmann scalar)
\be 
\rsq \equiv R^{\alpha \beta \gamma \delta}R_{\alpha \beta \gamma \delta}\,.
\ee
In the asymptotic limit  $\tau \to \infty$ with $v$ fixed, the general structure of this expression becomes
\be
\rsq =
\frac{N(v,s)}{\left( 1- \Delta^2 (s) v^8 \right) ^{4}}  + {\mathcal O} \left(\frac{1}{\tau^{\beta}} \right)\,,
\ee
for a positive $\beta$. The explicit expression of the numerator $N(v,s)$ is quite complicated and can be found in \cite{Janik:2005zt}. 
For generic $s$, the above asymptotic expression for $\rsq$ diverges at $v = \Delta^{-1/4}(s)$ and the singularity is a physical naked one. 
It turns out that only for $s = 4/3$ the fourth order pole of the denominator gets cancelled by an identical contribution in the numerator 
leaving a bounded scalar invariant. Only the physical singularity in the dual geometry of a perfect fluid (which will be found at $z=\infty$, see later on) can be hidden behind an 
event horizon. 

The above analysis suggests a criterion to select the profile $\epsilon(\tau)$ in (\ref{Tboost}). The proposal of \cite{Janik:2005zt} is that a physical boundary energy-momentum tensor has to 
be dual to a geometry satisfying the cosmic censorship. It then turns out that the requirement of the absence of naked singularities in the bulk uniquely selects perfect hydrodynamics 
as the large proper-time dynamics of the strongly coupled plasma. 

The scaling variable (\ref{scalingvariable}) and the energy density for a perfect relativistic fluid are
\be\label{v&epsilon}
v = \f{z}{\tau^{\f{1}{3}}}\,, \quad \epsilon(\tau)= \f{e_0}{\tau^{\f{4}{3}}}\,,
\ee
where we have reinstated a dimensionful parameter $e_0$ in $\epsilon$. 
In this case, the metric coefficients simplify considerably and the asymptotic geometry can be written in the compact Fefferman-Graham form
\bea
z^2 ds^2 &=& -\,  \f{\left( 1-\f{e_0}{3} \f{z^4}{\tau^{4/3}}\right)^2}{1\!+\f{e_0}{3}\f{z^4}{\tau^{4/3}}} \,\, d\tau^2 +dz^2\\
&& + \left(1+ \f{e_0}{3} \f{z^4}{\tau^{4/3}}\right) \left(\tau^2 dy^2  + dx^2_\perp \right)\,. \nonumber 
\label{movingBH}\eea 
This geometry is analogous to the static AdS black brane solution of section \ref{HR} and thus the potential singularity 
is indeed a coordinate one.  Here, however, the horizon position depends on proper-time 
\be
z_0= \left(\f{3}{e_0} \right)^{\f{1}{4}} \tau^{\f{1}{3}}\,.
\la{z0}\ee
Although a precise notion of temperature and entropy still lacks in a dynamical setup, in order to obtain a qualitative estimate one could
 naively generalize the static formula to infer 
\be
T =  \f{\sqrt{2}}{\pi z_0} = \left(\f{e_0}{3}\right)^{\f{1}{4}} \f{\sqrt{2} }{\pi \tau^{\f{1}{3}}}
\la{temp}\ee
for the temperature and the total entropy $S \propto s\tau= const.,$ see (\ref{entropy}). Observe that these estimates are in agreement with the ``Bjorken flow" 
expressions of equation (\ref{BjorkenFlow}).  Since the temperature of the black hole coincides with the one of the plasma, 
the gravity dual of the plasma cooling off during expansion is in terms of a black hole which moves off in the radial direction of the bulk, 
from the boundary towards the interior. A similar picture had already been qualitatively proposed in \cite{Shuryak:2005ia}.


\subsection{In-flow viscosity and relaxation time}\label{Viscosity&Relaxation}

It should be stressed that the results of section \ref{PerfectFluid} were obtained in the scaling limit. The gravity dual selects perfect hydrodynamics 
only in the strict asymptotic limit of infinite proper-time. Indeed, as shown for 
instance in \cite{Heller:2008fg} and \cite{Janik:2010we}, it is straightforward to repeat the computation of the square of the Riemann tensor up 
to the first subleading corrections in the metric and check that it is always singular for perfect fluid dynamics. The evaluation of $\rsq$ leads to
\be
\rsq = R_0(v) + \f{1}{\tau^{\f{4}{3}}}R_2(v)\,,
\ee
where $R_0(v)$ is finite, but $R_2(v)$ develops a $4^{th}$ order singularity. 
From the gravity side, it is then clear that the energy density in (\ref{v&epsilon}) needs to be corrected. 
It turns out that these corrections are due to viscous hydrodynamics, 
but it is instructive to see how the result arises using purely AdS/CFT methods. Assume that
\be
\epsilon(\tau) = \f{1}{\tau^{\f{4}{3}}} \left( 1- \f{2 \eta_0}{\tau^r} \right)
\ee
for a generic positive exponent $r$. Solving Einstein's equations with this boundary condition and computing the Kretschmann scalar yields a result of the form
\bea
\rsq = R_0(v) + \f{1}{\tau^r} R_1(v) &+& \f{1}{\tau^{2r}} \tilde R_2(v) \nonumber\\
&+& \f{1}{\tau^{\f{4}{3}}}R_2(v)\,,
\eea
where the last two terms are always singular. To obtain a bounded Kretschmann scalar, the only possibility is to make those two terms proportional to each other 
and therefore to set $r = 2/3$. This is exactly the scaling of a viscosity correction to Bjorken hydrodynamics with shear viscosity $\eta = \eta_0 / \tau$ 
(which follows from $\eta \propto T^3$ in the $\mathcal{ N} =4$ SYM case). Moreover, in \cite{Janik:2006ft} it has been shown that, 
to implement the cancellation of the potentially singular terms,  the coefficient $\eta_0$ has to be tuned to $\eta_0 = 2^{-1/2}3^{-3/4}$. 
This specific value corresponds to a shear viscosity to entropy ratio
\be
\f{\eta}{s} = \f{1}{4\pi}\,.
\ee
It is remarkable that the above result, obtained in a non-linear and dynamical setup coincides with the shear viscosity value obtained 
in \cite{son}, studying the response of a static plasma at fixed temperature to small perturbations. 

The procedure can be further generalized to higher orders to compute the other hydrodynamics coefficients. In general, the metric components
 $a(\tau,z)$, $b(\tau,z)$ and $c(\tau,z)$ in (\ref{FGansatz}) have an expansion of the form
\be
a(\tau,z) = \sum_n a_n(v) \f{1}{\tau^{\f{2n}{3}}}\,,
\ee
while the one of the curvature invariant is
\be
\rsq= \sum_n R_n \f{1}{\tau^{\f{4n}{3}}}\,.
\ee
The $2^{nd}$ order formalism is obtained truncating the sums above at $n=3$ and has been worked out in \cite{Heller:2007qt} (with a correction given in Ref.\cite{baier}, see further on). 
It results into a relaxation time 
\be
\tau_r = \f{2 - \ln 2}{2 \pi T}\,,
\la{time}\ee
which is about thirty times shorter than the one estimated on the basis of Boltzmann's kinetic theory. The value (\ref{time}) has been corrected by matching the results for the Bjorken flow of Ref.\cite{Heller:2007qt} with the full second-order formalism for hydrodynamics in conformal field theories at finite temperature derived in \cite{baier}.

Surprisingly, at this order, a new feature arises, namely the presence of a leftover logarithmic singularity in the expansion of $\rsq$. In \cite{Heller:2007qt}, 
it has been argued that such a divergence could be cured considering the coupling of a dilaton to the bulk metric. A non vanishing dilaton profile turns on, 
on the gauge theory side, a non-zero expectation value of Tr$F^2$, meaning that electric and magnetic modes are no longer equilibrated. The tuning of the 
dilaton field to achieve a non-singular Kretschmann scalar, leads to a negative expectation value $\langle$Tr$F^2\rangle < 0$, implying that electric 
modes dominate. 

However, the logarithmic divergences that arise at NNLO in the energy density expansion, have been considered more extensively in \cite{Benincasa:2007tp} 
and \cite{Heller:2009zz} (see also \cite{Kinoshita:2008dq} and \cite{Kinoshita:2009dx}). The conclusion is that 
the logarithmic singularity origins in a pathology of the Fefferman-Graham coordinates. There exists in fact a singular transformation to 
Eddington-Finkelstein coordinates, which yields a regular and smooth metric from the boundary up to the standard black brane singularity. In this coordinate system, 
the late proper-time expansions of curvature invariants is regular at the horizon. In order to proceed to higher orders, one would then have to perform the 
analysis in Eddington-Finkelstein coordinates. 
    

\subsection{Duality beyond boost-invariance}\label{BeyondBoostInv}

The efficacy of the approach that we presented in the previous sections resides in that no assumption about the 
dynamics of the plasma has to be made. It is the gravity dual analysis alone that selects the physically relevant evolution. 
However, in order to solve the full system of Einstein's equations one is in general constrained to work in highly symmetric 
setups and in specific regimes of interest. In particular, it is very hard to relax the assumption of boost invariance of the gauge theory matter. 
Significant progress in this direction has been achieved through a different approach in \cite{Bhattacharyya:2008jc}, 
where it has been shown in general how hydrodynamics arises from the gravity side.

The starting point of the analysis is the five-dimensional static AdS black brane which is dual to a strongly coupled plasma at rest 
($i.e.$ with flow vector $u^\mu = (1,0,0,0)$) and at temperature $T$. 
In incoming Eddington-Finkelstein coordinates, the metric reads 
\bea
ds^2&=& - r^2 \left[ 1- \left( \f{T}{\pi r}\right)^4\right] dv^2 + 2 \, dv dr\nonumber\\
&& + r^2 \eta_{i j} \, dx^i dx^j\,,
\eea
where the coordinate $r$ is related to the radial coordinate $z$ by $r = 1/z$. The conformal boundary of space-time is at $r = \infty$ 
and the horizon at $r = T/\pi$. 
This choice of the coordinate system is consistent with the comments of the end of section \ref{Viscosity&Relaxation}; these coordinates are well defined 
on the horizon and extend all the way from the boundary to the black brane singularity. 

By performing a boost, one can obtain the dual geometry to a uniformly moving plasma with $4$-velocity $u^\mu$
\bea\label{boostedBB}
ds^2& =& - r^2 \left[ 1- \left( \f{T}{\pi r}\right)^4\right] u_\mu u_\nu \, dx^\mu dx^\nu  \\
&&- 2 u_\mu \, dx^\mu dr  + r^2 (\eta_{\mu \nu} +u_\mu u_\nu )dx^\mu dx^\nu\,.\nonumber
\eea
The main idea of \cite{Bhattacharyya:2008jc} is to promote $T$ and $u^\mu$ to be slowly varying functions of the boundary Minkowski coordinates. In this way, 
the geometry ceases to be an exact solution of Einstein's equations because of the appearance of gradients of the temperature and of the velocity. The 
metric (\ref{boostedBB}) can be corrected order by order in a derivative expansion. The integration constants that arise at each order can be 
fixed requiring regularity of the metric at the horizon. The energy-momentum tensor of the plasma can then be read off from the corrected geometry and 
it is expressed, as the metric itself, in terms of $T$, $u^\mu$ and their derivatives. 
Explicit results have been worked out up to second order and read
\bea
T^{\mu\nu}&=&\f{1}{16 \pi G_5} \Bigg[ (\pi T)^4 (\eta^{\mu\nu}+4u^\mu u^\nu) \\ 
&& - 2(\pi T)^3 \sigma^{\mu\nu} +  (\pi T^2)  \Bigg( \log 2 \, T^{\mu\nu}_{2a}  \nonumber \\ 
&& +(2\!-\!\log2) \left( \f{1}{3}\, T^{\mu\nu}_{2c}\!+\!T^{\mu\nu}_{2d}\!+\!T^{\mu\nu}_{2e} \right) \Bigg) \Bigg]\,, \nonumber
\eea
where $G_5$ is the five dimensional Newton's constant and the rather lengthy expressions of the quantities $\sigma^{\mu \nu}$, $T_{2a}^{\mu\nu}$, etc. 
can be found in the original paper \cite{Bhattacharyya:2008jc}. 
The first term in $T^{\mu \nu}$ is simply the perfect fluid energy-momentum tensor, the second term involves the shear viscosity, while the third 
is the contribution of second order hydrodynamics. 

The conservation equation $\partial_\mu T^{\mu \nu} =0$ is equivalent, by definition, 
to the relativistic Navier-Stokes equation. Therefore, the dual geometry of every solution of the hydrodynamic equations is given by (\ref{boostedBB}) 
plus correction terms up to the same order in the gradient expansion.

We should note, however, that this construction is adequate exclusively for near the equilibrium setups. Starting from the boosted black brane metric, in fact, 
one implicitly assumes the existence of a hydrodynamical description, in terms of flow velocity and energy density. 
As we explain in section \ref{EarlyTime}, when dealing with processes that do not admit such a description, as the early stages of heavy-ion collisions, 
one has to recur to different methods. 


\section{AdS/CFT and early time quark-gluon plasma flow}\label{EarlyTime}

One of the main theoretical challenges of the physics of quark-gluon plasma is to describe its thermalization process.
 
The plasma is created in the initial stages of heavy-ion collisions, during which the system is certainly out-of-equilibrium. However, 
simulations based on the RHIC data suggest that an hydrodynamical description becomes reliable at times $< 1$ fm/c after the 
collision event \cite{Heinz:2004pj,Arnold:2004ti,Adcox:2004mh}. 
Perturbative QCD calculations \cite{Arnold:2004ti,Baier:2000sb,Molnar:2001ux} lead to a much longer thermalization time $\tau_{therm} \gtrsim 2.5$ fm/c, pointing 
in the direction of a strongly coupled scenario. As we saw, holographic methods are, up to now, one of the few tools of investigation available in this context, as 
well as a very promising one. 

In this section, we describe some steps forwards towards a description of the thermalization process of QGP, modeled by 
a strongly coupled $\mathcal N=4$ SYM plasma at finite temperature. Since it is a developing subject on a settled theoretical question dealing with far-from-equilibrium dynamics at strong coupling, we will discuss three different approaches.

i) In section \ref{response}, we presents estimates of the thermalization response-time computed through the quasi-normal modes of small perturbations 
away from the thermal background. This explores the information one can get on thermalization and isotropization from  {\it linearized} solutions, coupling non-hydrodynamic perturbations to a moving black hole background.

ii) Section \ref{Thermalization} is devoted to the discussion of the fully {\it non-linear} thermalization dynamics of a boost-invariant plasma, evolving 
from an anisotropic out-of-equilibrium early stage towards the hydrodynamic regime. The results point towards a dependence on initial conditions but also to the indication of a fast (if not complete) isotropization as 
characteristic of strongly coupled dynamics. 

iii) In section \ref{ShockWaves}  we mention the results 
obtained in modeling the Lorentz-contracted relativistic nuclei of a heavy-ion collision through initial shock waves and look for a solution of the dual gravitational field after the collision. This provides a realistic scenario, at least from the kinematic point of view. 

A warning is in order to remind the following unsolved problem. The preferred phenomenological scenario for the QGP formation starts from initial conditions which may be dominated by weak QCD coupling, due to the high density of partons in a fast boosted nucleus. 
Hence the problem remains open to describe on a microscopic basis the transition from a far-from-equilibrium weakly coupled system towards a hydrodynamic one at strong coupling. At least the AdS/CFT correspondence allows to make a few first (and quite unique at present) steps on the understanding of strongly coupled dynamics.


\subsection{Thermalization response-time}\label{response}

\subsubsection{Quasi-normal modes}

The quasi-normal modes of a black hole describe the response of the system after a small perturbation.  Since the field that sources the perturbation can fall into the black hole or 
radiate to infinity, the modes of oscillations decay, their frequencies being complex. The geometry therefore undergoes damped oscillations that produce an exponential decay of the 
perturbation. The frequencies and damping of these oscillations are entirely fixed by the black hole, and are independent of the initial perturbation. 
For black holes in asymptotically flat space-times, the exponential decay of the quasi-normal modes is followed by a power-law decrease \cite{Price:1971fb}.
The decay is instead purely exponential in the case of black holes in AdS \cite{Ching:1995tj}. 

In the AdS/CFT framework, perturbing a static black hole in AdS corresponds to perturbing the approximately thermal 
dual state, and the decay of the perturbations describes relaxation back to thermal equilibrium. Computing the quasi-normal modes of the equations of motion linearized around the background  
allows therefore to estimate the thermalization response-time of the strongly coupled dual gauge theory after a small deviation from equilibrium. 

The behavior of quasi-normal modes should be contrasted with the one of hydrodynamical excitations that go to zero at small transverse momentum and have a slower power-law decay. 
In the static case, for instance, both quasi-normal modes and viscosity calculation correspond to poles of specific retarded propagators, respectively for a scalar and metric deformation, through Fourier transform in the time 
variable. However, in the limit of zero transverse momentum, the poles corresponding to the viscosity case vanish. The ``non-hydrodynamic'' nature of the scalar quasi-normal modes allows instead 
to obtain a finite thermalization response-time in this limit. 

The quasi-normal modes analysis has been carried out in \cite{Horowitz:1999jd} 
for black holes in four, five and seven dimensional global AdS and perturbations produced by a minimally coupled scalar field excitation. See also \cite{Starinets:2002br,Friess:2006kw,Iqbal:2008by} for 
later developments. In asymptotically AdS space times, quasi-normal modes are defined as solutions of the wave equation that are purely ingoing near 
the horizon and vanish at the boundary. For a minimally coupled scalar perturbation, they are obtained solving the wave equation for a massless scalar field $\nabla^2 \Phi =0$ in the gravitational background, 
with absorbing boundary conditions at the horizon $z=z_0$ and Dirichlet conditions at the boundary $z\!=\!0$. One writes 
\be
{{{\nabla^2 \Phi \equiv \f{1}{\sqrt{-g}} 
\partial_\al 
\left(\sqrt{-g} g^{\al\bt}\partial_\bt \Phi \right)=0}}}
\la{wave}\ee
for the standard covariant coupling of the scalar field $\Phi$ to the background metric $g^{\al\bt}$ of determinant $g$.

\subsubsection{Static black hole}

Since the main example throughout the notes is $\mathcal N=4$ SYM in four-dimensional Minkowski space, 
in the following we refer the results obtained in the case of a static planar black hole in AdS$_5$. To facilitate the comparison with the results of \cite{Janik:2006gp} discussed later in the section, 
we work in Fefferman-Graham coordinates. 

Recall that in these coordinates, see (\ref{e.bhfef}), the metric of a five-dimensional black brane reads
\bea\label{5dBH}
ds^2=-\f{\left(1\!-\!\f{z^4}{z_0^4}\right)^2}{\left(1\!+\!\f{z^4}{z_0^4}\right)} \f{dt^2}{z^2}\! +\! \left(1\!+\!\f{z^4}{z_0^4}\right) \f{d\vec x^2}{z^2}\! + \! \f{dz^2}{z^2}.
\eea
where $(t, \vec x)$ are the boundary coordinates, $z$ is the AdS radius and $z_0$ is the location of the event horizon in the bulk.

In the case of the static metric (\ref{5dBH}), Eq.(\ref{wave}) writes
 \bea
&-&\f{1}{z^3} 
\f{\left(1-\f{z^4}{z_0^4}\right)^2}{1+\f{z^4}{z_0^4}}\ \partial_t^2 \Phi(t,z)\nonumber\\
&+&\partial_z \left[\f{1}{z^3} \left(1-\f{z^8}{z_0^8}\right) \partial_z
\Phi(t,z) \right]=0\ .
\la{wavestat}\eea
Using the separation of variables 
\be
\Phi(t,z) = e^{i \omega t}\ \phi(z)\,,
\ee
the wave equation (\ref{wavestat}) leads to a Heun equation in the bulk variable \cite{Starinets:2002br} 
\bea \label{Heun}
\phi^{\prime\prime} &+& \frac{1 - \tilde z^2}{\tilde z (1-\tilde z)(2-\tilde z)} \phi^{\prime}  \nonumber\\
&& +\frac{\left({z_0\omega}\right)^2}{8\tilde z (1-\tilde z)(2-\tilde z)} \phi =0 \nonumber \\
&&
\eea
 where $\tilde z \equiv (1 - (z/z_0)^2)^2 / (1 +(z/z_0)^4)$, $z_0={\sqrt 2}(\pi T)^{-1},\ cf.$ (\ref{haw}), and {the prime denotes the derivative with respect to $\tilde z$}. 
 
The quasi-normal modes analysis then proceeds through a combination of analytic and numerical methods (see \cite{Horowitz:1999jd} for details). 
Note that quasi-normal modes appear also as poles in the complex frequency plane of the Fourier transform in time $t$ of the retarded Green's function (see $e.g.$ \cite{sonsta}). 

Interestingly enough, for a homogeneous scalar excitation, the dominant decay mode at large time is given by a quasi-normal mode whose 
frequency acquires a non zero imaginary part and thus corresponds to a minimal exponential decay mode with temperature. Technically, by matching  Eq.(\ref{Heun}) with a Shr\"odinger equation
\be
\partial^2_{z^*} \phi(z^*) 
+\ \f{\sqrt{8}\ e^{12z^*}}{\sinh^{3/2}(8z^*)}{\left(\f{\om}{\pi  
T}\right)^2  \phi(z^*)}=0
\la{"}\ee
 through the change of variable $$z\to z^*\equiv \f 14\ {[\tanh^{-1}(z^4/z_0^4)]^4},$$ one gets quantized  ``energy states''.  The  dominant exponential decay mode, corresponds to the state with minimal energy in the Shr\"odinger equation (\ref{"}).
Denoting $\omega_k$ the eigenvalues of Eq. (\ref{"}), one finds
\be 
\f{\omega_k}{\pi T} = a_k\! -\!ib_k\,,
\ee
and
\be \label{dominantQNM}
\cdots>b_k>\cdots >b_1=2.74667 \,,
\ee
where $T$ denotes both the plasma and black brane temperature. The corresponding response timescale is thus given by the first eigenvalue $\om_1$ and leads to $\tau_{resp} = 1/ \Im m\omega_1 \sim 0.116 /T$.

\subsubsection{Moving black hole}

A similar analysis was carried out in \cite{Janik:2006gp} for a black hole moving away in the radial bulk direction. This geometry was discussed in section 
\ref{PerfectFluid} and is the gravity dual of a boost-invariant expanding perfect fluid in $\mathcal N=4$ SYM at large proper-time. Knowledge of the quasi-normal 
modes of this black hole  gives an estimate of the {\it thermalization response-time} of the strongly coupled gauge theory, that is the relaxation time after a perturbation of the local thermal equilibrium due to the coupling to a scalar field. 

The background geometry describing a black hole moving off from the boundary in the $z$ direction was given in (\ref{movingBH}). In terms of the scaling variable 
$v= z \tau^{-1/3}$, it writes
\bea \label{movingBH2}
z^2 ds^2& =& -\frac{(1-v^4)^2}{(1+v^4)}d\tau^2 +  \nonumber \\
&& + (1+v^4)(\tau^2 dy^2 + dx_{\perp}^2)+ dz^2\,.
\eea
Consider now the wave equation of a canonically coupled scalar field $\Phi$ in the boost-invariant setting. 
In the scaling limit $v=$const and $\tau \to \infty$, the corresponding Klein-Gordon equation (\ref{wave}) is now expressed in terms of the two variables $\tau$ and $z$. 
Through the change of variables $(\tau ,z)\to(\tau, v)$,
so that 
\be
\partial_z \to \tau^{-\f{1}{3}} \partial_v\,, \quad  \partial_\tau \to \partial_\tau -\f{1}{3}\tau^{-\f{4}{3}} \partial_v \,,
\ee
and neglecting the non diagonal terms in the large $\tau$ expansion, one gets
 \bea
&-&\f{1}{v^3} 
 \f{(1+v^4)^2}{1-v^4} \partial_\tau^2 \Phi(\tau,v)\nonumber\\
&+&\tau^{-\f{2}{3}}\ \partial_v \left(\f{1}{v^3} (1-v^8) \partial_v
\Phi(\tau,v) \right)=0\ ,
\la{wavemov}
\eea
which is similar to (\ref{wavestat}) but with a $\tau$-dependent coefficient. 
Performing the separation of variables 
\be \label{separation}
\Phi(\tau ,z) = f(\tau) \phi(v)\,,
\ee 
it reduces to two decoupled equations 
\bea
\ \ \ \partial_\tau^2 f(\tau)  + \hat\omega^2 \tau^{-\frac 2 3} f(\tau)\!\!\!\!\!\!\!\!& &= 0  \label{QNMs1}\\ \!\!\!\!\!\!\!\!\!\!\!\!\!\!\!\!\!\!\!\!\!\!\!\!
\partial_v \left[ \frac{(1\!-\!v^8)}{v^3} \partial_v \phi(v) \right]\!\!\!\! &+&\!\!\! \!\hat\omega^2 \frac{(1-v^4)^2}{v^3(1\!+\!v^4)}\ \phi(v)=0\,. \nonumber\\
&& \label{QNMs2}
\eea
where we introduced an {\it a priori} arbitrary parameter $\hat\omega,$ whose allowed values will be fixed by the (quantized) solutions of (\ref{QNMs2}).

Equation (\ref{QNMs1}) is solved by linear combinations of the Bessel functions
\be
 \sqrt{\tau} J_{\pm \frac 3 4} \left( \frac 3 2 \hat\omega \tau^{\frac 2 3}\right)\,, 
\ee
with relevant large $\tau$ behavior
\be \label{taudependence}
 f(\tau) \sim \tau^{\frac 1 6} e^{\frac 3 2 i({\hat \om_1}) \tau^{\frac 2 3}}
\ee
where $\Im m\  \hat \om_1=2.74667$ is the same as in (\ref{dominantQNM}). Indeed,
the equation in the $v$ variable (\ref{QNMs2}) is formally identical to the Heun equation (\ref{Heun}) of the static black hole and therefore leads to the same numerical values for the
quasi-normal modes and frequencies. However, the variables $v$ and $\tau$ of (\ref{separation}) are  different from the variables $z$ and $t$ relevant in the static case. 
Moreover, the proper-time dependence (\ref{taudependence}) has a non trivial scaling compared to the plane wave dependence of the static solution. 

Noting that the moving  horizon is at $
z_0= \left(\f{3}{e_0} \right)^{\f{1}{4}} \tau^{\f{1}{3}}=\f {\sqrt 2}{\pi T},\ cf.$ (\ref{z0},\ref{temp}),
it was suggested  in \cite{Janik:2006gp}, that this result could be understood in terms of an adiabatic approximation of the expanding case, with locally fixed temperature. Indeed, 
\be
{\vert f(\tau)\vert \sim e^{-\f{3}{2} \Im m ({\hat \om_1})\ \tau^{\f{2}{3}}} \sim e^{-\pi b_1 T\tau}}\ ,
\ee  
where $\om_1=\pi T(a_1-ib_1)$ is the dominant decay mode in the static case (\ref{dominantQNM}).

Plugging in the numerical value of the dominant decay mode 
(\ref{dominantQNM}) in (\ref{taudependence}) leads to a damping factor of the form
\be
\exp \left( - \frac 3 2 \cdot 2.7466 \cdot \tau^{\frac 2 3}\right)\sim  e^{-8.3 T\tau}\ .
\ee
The above results were extended in \cite{Friess:2006kw} to include also vector and tensor quasi-normal modes. 

In the limits to which this analysis can be considered quantitatively relevant for the thermalization in heavy-ion reactions, the fast decay modes of frequency (\ref{dominantQNM})
damp out to $1/e$ of their original amplitude in a time no greater than 
\be
\tau_{resp} \Big|_{T= T_{peak}} = \frac{0.116}{T} \Big|_{T= T_{peak}} \approx 0.08 \textrm{ fm/c}\,,
\ee
for a typical initial peak gauge theory temperature $T_{peak} \approx 300$ MeV, which is commonly considered as reasonable in phenomenology. Given the highly anisotropic momentum space distribution expected in the early stages of 
a RHIC collision, one tentatively estimates a certain finite number $n$ of  e-foldings of the ``response thermalization process'' have to elapse before hydrodynamic approximations 
can be used. The estimated  thermalization time becomes
\be
\tau_{therm} \sim n \times \tau_{resp} \Big|_{T= T_{peak}} \,,
\ee
leading to  $\tau_{therm}\approx 0.3 \textrm{ fm/c}$ following  the guess of \cite{Friess:2006kw} of about $n\!=\!4$ iterations. The above estimate compares in order of magnitude  with the simulations predictions on real experiments. Nevertheless, it should be clear that it is extrapolated from 
the quasi-normal modes analysis, that, by definition, only applies to the late-time stages of the whole thermalization process. Moreover, the thermalization response time has been obtained in a linearized approximation around the black hole background through AdS/CFT methods. 

As a final remark on this response time studies, it is worth noting that the same properties are valid for  quasi-normal modes dominating the transverse space diagonalization of the metric after a non diagonal perturbation. Let us indeed introduce a  non diagonal transverse component $g_{x_1,x_2}$ using its contravariant expression 
\be
g^{x_1}_{x_2}(z,\tau ) \equiv \f {z^2}{1+v^4}  g_{x_1x_2}\ . 
\ee
It  verifies (with $g^{x_1}_{x_2}(z,\tau)=f(\tau)\ g^{x_1}_{x_2}(v)$) the same equations (\ref{QNMs1},\ref{QNMs2}); such as 
\bea
\partial_v \left(\f{1}{v^3} (1-v^8) \partial_v g^{x_1}_{x_2}(v) \right)\! \! 
&+&\!\!  \f{\hat \om^2}
{v^3} \f{(1+v^4)^2}{1-v^4} g^{x_1}_{x_2}(v)\nonumber\\&=&0\ .
\eea
In some sense, isotropization  and thermalization response-times are equal at large proper-times.  We now will turn to the study of the non-linear problem of isotropization and thermalization when starting from far-from-equilibrium 
initial conditions.


\subsection{Thermalization of boost-invariant plasma}\label{Thermalization}

In the present section we describe three different approaches to the study of the thermalization process of a strongly coupled boost-invariant plasma. 
The analysis are to be intended as complementary and point towards a fast isotropization/thermalization. However, as we shall discuss, this represents only an early stage for a full understanding of far-from-equilibrium dynamics
leading to the QGP formation. 


\subsubsection{Solution with ``full anisotropy''}

One of the first holographic estimates of the isotropization/thermalization time of the quark-gluon plasma appeared in \cite{Kovchegov:2007pq}. The setup of this analysis is similar to 
the one considered in \cite{Janik:2005zt}, but with differences which we will comment at the end of this subsection.

The starting proposal  of the paper is that Bjorken hydrodynamics 
at late times can be singled out, in the holographic renormalization program, relaxing the assumptions to only require that the metric tensor is a real and single-valued function 
of the coordinates everywhere in the bulk, without imposing any constraint on the curvature invariants. Applying the same strategy to early time dynamics  \cite{Kovchegov:2007pq} 
leads to infer  the existence of a solution corresponding to the fully anisotropic case with constant energy density $\epsilon$. In the notation of section \ref{Boost-invFlow}, 
it corresponds to the solution with $s=0$ in equation (\ref{fboost}), and thus to ``full anisotropy'' at initial time, namely $\eps= p_\perp=-p_\parallel$. 
The system is initially anisotropic, with negative longitudinal pressure, and is expected to evolve at late times to ideal Bjorken hydrodynamics, with longitudinal and transverse pressure 
components equal and positive. Isotropization must take place at some intermediate proper-time. 

Assuming a smooth transition between full anisotropy and full isotropy, the isotropization time $\tau_{iso}$ can be estimated matching the small and large proper-time regimes. In \cite{Kovchegov:2007pq}, this proper-time was defined as the crossing value 
of the branch-point singularities of the two regimes, leading to
\be\label{tauiso}
\tau_{iso} = \left( \frac{3 N_c^2}{2 \pi^2 e_0} \right)^{\frac 3 8}\,.
\ee  
 Here $e_0$ is the same dimensionfull parameter that appeared in (\ref{v&epsilon}) in the late proper-time behavior $\epsilon(\tau) = e_0 \tau^{-4/3}$.
 
Phenomenologically, to obtain a rough estimate of the isotropization time (\ref{tauiso}), one can extrapolate the realistic evaluations of RHIC to the supersymmetric case under consideration. 
The hydrodynamical simulations for central $Au +Au $ collisions with $\sqrt{s} = 200$ GeV yield the energy density $\epsilon = 15$ GeV/fm$^3$ at the proper-time $\tau = 0.6$ fm/c. 
Plugging in the corresponding value of $e_0$ in (\ref{tauiso}) and setting $N\!=\!3$, in  \cite{Kovchegov:2007pq} it was obtained $\tau_{iso} \approx 0.3$ fm/c. 

Ref.\cite{Kovchegov:2007pq}  thus points towards a small value of the thermalization time. However, the analysis relies on the assumption of a rapid and smooth interpolation between the initial fluid conditions and  the late time regime. 
A full solution of that transition remains necessary, as we shall discuss now, either by starting from given initial conditions or by a general study of the non-linear gravitational solution in the bulk of the AdS space.


\subsubsection{Far-from-equilibrium forcing dynamics}\label{CYForcingDynamics}

A way to prepare an out-of-equilibrium state is to turn on, in the ground state of a system, a time-dependent perturbation of the background fields. 
For instance, one can introduce a time-dependent deformation of the boundary geometry with compact support. From an holographic perspective, this acts as a source of gravitational 
radiation in the bulk, leading to gravitational collapse and inevitably to black hole formation. After the perturbation is turned off, the geometry relaxes to a 
smooth and slowly varying form. In the dual theory, the latter process describes relaxation of the non-hydrodynamic degrees of freedom towards perfect fluid-dynamics. 

This approach has been taken in \cite{Chesler:2009cy} (following previous analysis \cite{Chesler:2008hg}) to study the thermalization process of a boost-invariant $\mathcal N=4$ SYM plasma. 

The familiar Ansatz of two-dimensional spatial homogeneity, $O(2)$ rotational invariance in the transverse plane (defined by $x_{\perp}$) and boost-invariance in the 
longitudinal direction ($x_{\parallel} = x_1$) has been considered in \cite{Chesler:2009cy}. This setup has the advantage of allowing the comparison with the late proper-time 
results of \cite{Janik:2005zt}. Since the late hydrodynamic evolution is nearly isotropic, it is especially interesting to generate a high anisotropy in the initial non-equilibrium state.

\begin{figure}[t]
\begin{center}
\includegraphics[scale=0.25]{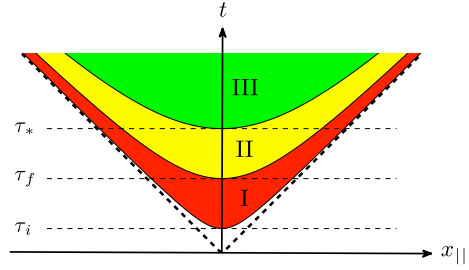}
\caption{Space-time diagram of the field theory plasma evolution. The metric deformation is turned on in the time interval $(\tau_i, \tau_f)$. 
At $\tau_*$ hydrodynamics sets in. Figure  from \cite{Chesler:2009cy}.}
\label{fig:spacetime}
\end{center}
\end{figure}

To get some intuition on the process, we borrow Fig.\ref{fig:spacetime} from \cite{Chesler:2009cy}. The figure depicts a space-time diagram of the field theory evolution. 
The system starts out at proper-time $\tau =0$ in the ground state. At $\tau = \tau_i$, the four-dimensional boundary geometry starts to deform in the vicinity of $x_{\parallel} =0$ and the perturbation acts until time $\tau_f$, 
propagating in the $\pm x_{\parallel}$ directions at speeds asymptotically approaching the speed of light. 
The relevant space-time region is I in red. At $\tau =\tau_f$, the field theory is out-of-equilibrium and significantly anisotropic. It relaxes towards local equilibrium and hydrodynamics in region II in 
yellow. After time $\tau_*$, region III in green, the system is well approximated by a perfect fluid behavior. Deriving the thermalization time of the plasma amounts to the computation of the 
proper-time $\tau_*$ in the diagram.  

In \cite{Chesler:2009cy}, the deformation of the boundary geometry was taken to be produced by a time-dependent shear
\be \label{metricshear}
d s^2 = - d\tau^2 + e^{\gamma(\tau)} d \vec x_{\perp} + \tau^2 e^{-2 \gamma(\tau)} dy^2\,.
\ee
Here
\be
\gamma(\tau) = c \, \Theta \left[\delta(\tau) \right] \delta(\tau)^6 e^{-\frac{1}{\delta(\tau)}}\,,
\ee
with
\be
\delta(\tau) = 1- \frac{(\tau - \tau_0)^2}{\Delta^2}\,;
\ee
$c$ is a constant characterizing the amplitude of the perturbation and $\Theta$ is the unit step function. The deformation acts during the time interval $(\tau_i, \tau_f)$, 
with endpoints $\tau_i \equiv \tau_0 - \Delta$ and $\tau_f \equiv \tau_0 + \Delta$. Choosing $\tau_0 \equiv 5/4\, \Delta$, the geometry is flat at $\tau =0$. All quantities will be 
measured in units where $\Delta =1$, so that $\tau_i = 1/4$ and $\tau_f=9/4$.

The most general bulk Ansatz consistent with diffeomorphism and spatial 3D translational invariance, together with $O(2)$ rotation invariance, is 
\bea
ds^2 = -A d\tau^2 &+& \Sigma^2 \left[ e^B d \vec x_{\perp}^2 + e^{-2B} dy^2 \right]\nonumber\\
 &+& 2 dr d \tau \,,
\eea
where $A$, $B$ and $\Sigma$ are all functions of the bulk radial coordinates $r$ and $\tau$ only. The coordinate $r$ is related to the previously used $z$, by $r = 1/z$. 
Notice that here $\tau$ and $r$ are generalized infalling Eddington-Finkelstein coordinates. 

Einstein's equations, subject to the boundary conditions that the boundary metric $g_{\mu\nu}(x)$ coincides with (\ref{metricshear}), can be solved numerically. The boundary stress-energy 
tensor can be computed through
\be
T^{\mu \nu}(x) = \frac{2}{\sqrt{-g(x)}} \frac{\delta S_G}{\delta g_{\mu \nu}}\,,
\ee
where $S_G$ denotes the gravitational action. The time-dependence of the energy density, longitudinal and transverse pressures then follows from (see \cite{Skenderis})
\be
T^\mu_\nu = \frac{N_c^2}{2\pi^2} \textrm{diag} (- \epsilon, p_{\parallel}, p_{\perp} ,p_{\perp})\,.
\ee
\begin{figure}[t]
\vspace{1.7cm}
\includegraphics[scale=0.2]{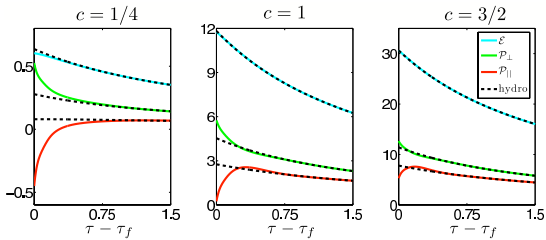}
\caption{Energy density (blue top curve), transverse pressure (green middle line) and longitudinal pressure (red bottom), divided by $N_c^2/2\pi^2$, as a function of time for $c=1/4$, $c=1$ and $c=3/2$. 
The dashed black lines show the second order viscous hydrodynamic approximation to the different stress tensor components.}
\label{fig:pressures}
\end{figure}
The numerical solution explicitly shows the formation of an event and apparent horizon, which build up in the bulk and merge at late times.
At asymptotically late times, the metric settles down to the moving black hole dual of the perfect fluid (\ref{movingBH}). 

The transition from the far-from-equilibrium behavior to near-local-equilibrium hydrodynamics is found to be regulated by the relative 
importance of the exponentially decaying non-hydrodynamic degrees of freedom in comparison to the slowly relaxing hydrodynamic modes. This is in contrast with the naive expectation that hydrodynamic should 
set in at the time at which higher order terms in the hydrodynamic expansion become comparable to lower order terms. 

A quantification of the results is provided in figure Fig.\ref{fig:pressures}, which was presented in \cite{Chesler:2009cy}. The energy density ($\epsilon$), transverse and longitudinal pressures ($p_{\perp}$ and $p_{\parallel}$) are plotted 
for different values of $c$, as a function of time, starting at $\tau =\tau_f$. On top of the numerical data are represented the hydrodynamic approximations. The plots show a significant anisotropy, even at late times, where hydrodynamics applies. 
For all values of $c$, the transverse pressure approaches the longitudinal pressure from above. This statement can be checked by perturbative computations of the first order viscous corrections. As the value of $c$ increases, the 
system gets closer to equilibrium at $\tau_f$. Indeed, for larger $c$, the perturbation does more work on the geometry and the system reaches a higher temperature. In a conformal theory, as $\mathcal N=4$ SYM is, the relaxation time of the non-hydrodynamic 
degrees of freedom must scale inversely with the temperature. For $c \to \infty$, the system will always be close to local equilibrium and isotropic already at $\tau_f$.  

The thermalization time $\tau_*$ was defined in \cite{Chesler:2009cy} as the time beyond which the stress tensor agrees with the hydrodynamic approximation within $10 \%$. 
In all the analyzed cases, the relevant dynamics, from the production of the plasma to his relaxation to near to local equilibrium, takes place over a time $\tau_* - \tau_i  \lesssim 2/T_*$, where $T_*$ 
denotes the local temperature at the onset of the hydrodynamic regime. 


\subsubsection{General approach to the early-time flow}\label{BoostEarlyTime}

In this section we discuss the results obtained in \cite{Beuf:2009cx}, where an analysis of the early time expansion of the boost-invariant flow has been 
carried out on the lines of \cite{Janik:2005zt}. We refer the reader back to section \ref{Boost-invFlow} for the setup and the basic assumptions.

The idea of the approach is the same as for the large proper-time regime discussed in sections \ref{LargeTime} and \ref{PerfectFluid}. 
In the following, we describe how to holographically reconstruct the dual geometry to a plasma configuration with stress-energy tensor (\ref{Tboost}) 
at early proper-times. This is accomplished solving Einstein's equations in a power series expansion starting from the conformal boundary of 
space-time and subject to the appropriate boundary conditions. 

In the late times case, the structure of Einstein's equations naturally led to the introduction of a scaling variable and to a remarkable 
simplification of the computations. 
It turns out that it is not possible in general to apply the same procedure to the small $\tau$ regime. In what follows 
we schematically reproduce the main steps of the computation and point out the complications encountered in \cite{Beuf:2009cx}.

We recall that the most general Ansatz for the bulk metric consistent with the plasma symmetries is
\bea\label{FGansatz2}
z^2 ds^2 &= &-e^{a(\tau,z)} d\tau^2 +\tau^2 e^{b(\tau,z)} dy^2\nonumber\\
&&  + \, e^{c(\tau,z)} d \vec x^2_{\perp} +dz^2\,.
\eea
Solving Einstein's equations (\ref{Einstein'seqs})
\be
R_{\alpha \beta} = \frac{1}{2}g_{\alpha \beta} (R -12) 
\ee
in a power series expansion in the bulk coordinate $z$, together with the boundary conditions (\ref{BCa})
\be
a(\tau,z) = - z^4 \epsilon(\tau) + z^6 a_6(\tau) + z^8 a_8(\tau) + \dots
\la{}\ee
allows to fix the metric components $a(\tau,z)$, $b(\tau,z)$ and $c(\tau,z)$ order by order in the $z$ expansion. 

Analyzing the structure of the early-time expansion of the equations, in \cite{Beuf:2009cx} it was concluded
that a scaling variable does not exist in general in the $\tau \to 0$ limit. As we shall see, the solution of \cite{Kovchegov:2007pq} is not unique. One has  to analyze the full solution at $\tau \sim 0$. 

From a physical point of view, it is natural to expect that 
the initial conditions should play a crucial role at early times. On the gauge theory side, the initial state can be prepared in a multitude of ways and therefore 
its early-time behavior cannot be expected to be universal. This contrasts the late time dynamics which, under the specific symmetry assumptions under consideration, 
is governed by a single temperature scale. The dissipative effects which act during the plasma evolution wash out the 
differences due to the initial conditions leading to a hierarchy of terms and to a single scale governing the large time expansion of the energy density.

In the absence of a scaling argument, one can still determine the small $\tau$ behavior of the energy density $\epsilon(\tau)$ assuming that at $\tau=0$ the  
initial condition is regular. This assumption does not need to be a priori relevant for realistic heavy-ion collisions, but it is in any case an interesting analysis 
in its own. It implies in fact a finite limit of $\epsilon(\tau)$ as $\tau \to 0$, consistent with the $s=0$ behavior found in \cite{Kovchegov:2007pq}. 
Moreover, it restricts the expansion of the energy density 
to be only in even powers of the proper-time
\be
\epsilon(\tau) = \epsilon_0 + \epsilon_2 \tau^2 + \epsilon_4 \tau^4 + \dots
\ee
The coefficients $\epsilon_{2n}$ are uniquely determined, through Einstein's equations, in terms of the coefficients of the initial condition for the metric
\be
a_0 (z) \equiv a(\tau =0, z) 
\ee
and similarly for the other components. 

The admissible initial conditions are given by the constraint equations contained in the full set of Einstein's equations. 
These imply $a_0(z) = b_0(z)$. Defining 
\bea
v(z^2)& \equiv &  \frac{1}{4z}a_0^{\prime}(z) =  \frac{1}{4z}b_0^{\prime}(z)\\
w(z^2) &\equiv &  \frac{1}{4z}c_0^{\prime}(z)\,,
\eea
the remaining constraint reads
\be \label{constraint}
v^{\prime} + w^{\prime} +v^2 +w^2 =0\,,
\ee
where the prime denotes the $z^2$-derivative. It is easy to see that there does not exist an anywhere bounded solution 
of the constraint equation. Integration of (\ref{constraint}) gives
\bea
0 &=& \int_0^{\infty} (v^{\prime} + w^{\prime} )  dz^2 +  \int_0^{\infty}(v^2 +w^2) dz^2 \nonumber \\
& =& \int_0^{\infty}(v^2 +w^2) dz^2\,,
\eea
since the first integral vanishes because of the boundary conditions. The only regular solution is vacuum AdS$_5$ with $v=w=0$. 
Therefore, at any time, including $\tau=0$, the metrics of interest are singular. 

The requirement imposed in \cite{Beuf:2009cx} of the absence of 
curvature singularities (other than the one at $z=\infty$) is thus a very powerful tool to select the allowed initial conditions. 
The coordinate singularity that is then left at all times in the bulk might signal the presence of a dynamical horizon, although this interesting statement requires 
further investigation. 

A bonus that follows from the simplicity of (\ref{constraint}) is that it can 
be solved analytically. One neat example of a solution to the initial value constraint, satisfying the non-singularity requirement, reads
\bea\label{initialcond}
a_0(z) &=& b_0(z) = 2 \ln \cos a z^2\,, \nonumber \\
c_0(z) &= & 2 \ln \cosh a z^2\,,
\eea
for a generic constant $a$. 
The details of a larger number of allowed initial conditions can be found in \cite{Beuf:2009cx}. Some of them are also mentioned in the following. 
 \begin{figure}[t]
\begin{center}
\includegraphics[scale=0.25]{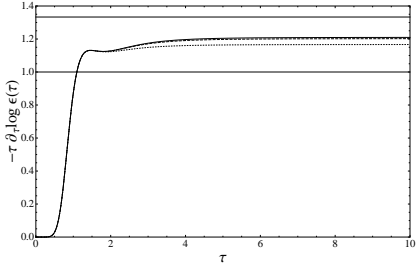}
\caption{Pad\'e approximation of the effective power $s_{eff}(\tau)$ in the energy density for expansions up to order $\tau^{64}$ (dotted line), $\tau^{80}$ (dashed line), 
$\tau^{96}$ (solid line) and initial conditions (\ref{initialcond}). The two horizontal lines denote $s=1$ (free streaming scenario) and $s=4/3$ (perfect fluid case).}
\label{fig:effectiveexp}
\end{center}
\end{figure}

Given the class of acceptable initial conditions, one then needs to solve Einstein's equations with these initial data to obtain the profile $\epsilon(\tau)$. 
In the absence of a scaling variable, the solution has to be exact and the ideal treatment of the equations should be numerical. 
This analysis is in progress \cite{Progress}. 
 
In a primary exam, an analytic analysis was pursued in \cite{Beuf:2009cx} where the equations were solved in a power series in $z$ and $\tau$ for 
specific initial conditions. The limit of this approach is that the power series for the energy density has a finite radius of convergence necessitating the use of 
as a resummation scheme, such as Pad\'e approximation. 

The system evolves from an early time dynamics governed by the initial conditions to an hydrodynamical regime at late times. One way to quantify this 
transition is through the exam of the effective exponent of the power-law dependence of the energy density
\be
s_{eff}(\tau) \equiv -\tau \frac{d}{d\tau} \ln \epsilon (\tau)\,.
\ee 
In Fig.\ref{fig:effectiveexp}, this effective power is plotted as a function of $\tau$ for the initial conditions (\ref{initialcond}). The plot is 
taken from \cite{Beuf:2009cx}. The different curves correspond 
to different cuts in the Pad\'e approximation. All profiles start out at zero, clearly cross the line $s=1$ denoting the free streaming scenario, and move upwards 
towards $s= 4/3$, which corresponds to the perfect fluid flow. However, to be sure that the profile indeed reaches $s=4/3$, a numerical solutions is required. 

\begin{figure}[t]
\centering
\includegraphics[height=0.8in,width=0.3\linewidth]{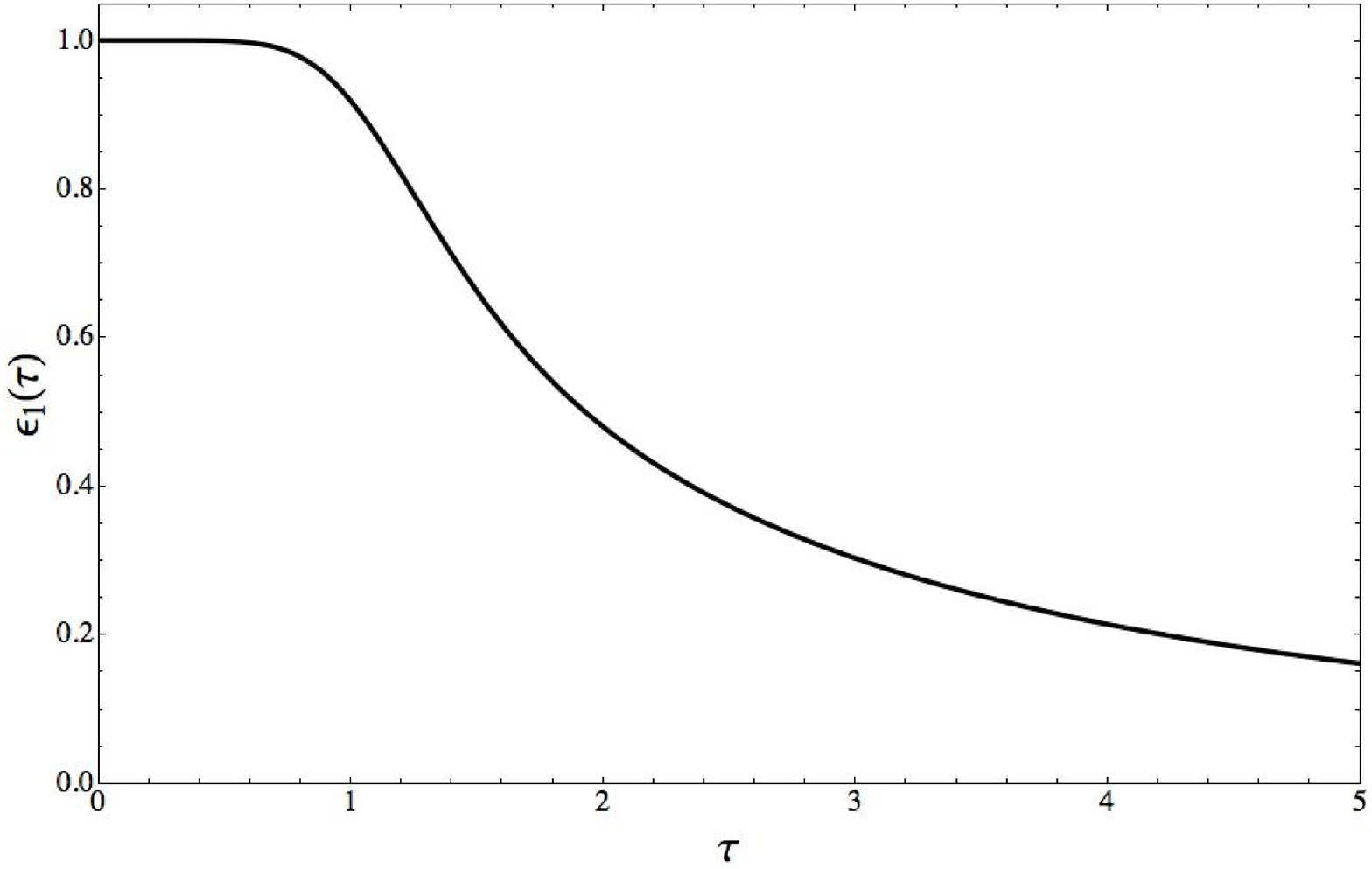}
\includegraphics[height=0.8in,width=0.3\linewidth]{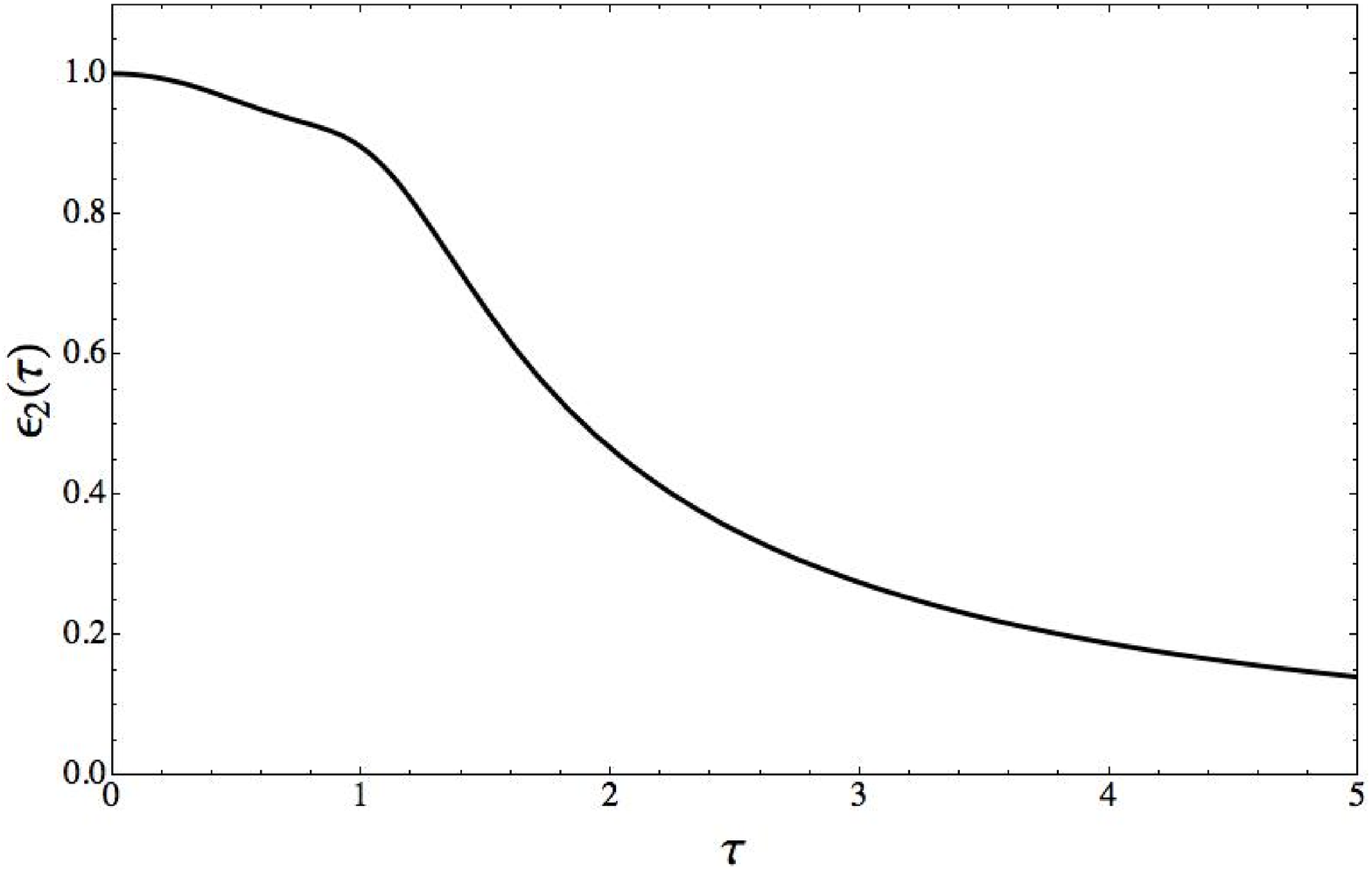}
\includegraphics[height=0.8in,width=0.3\linewidth]{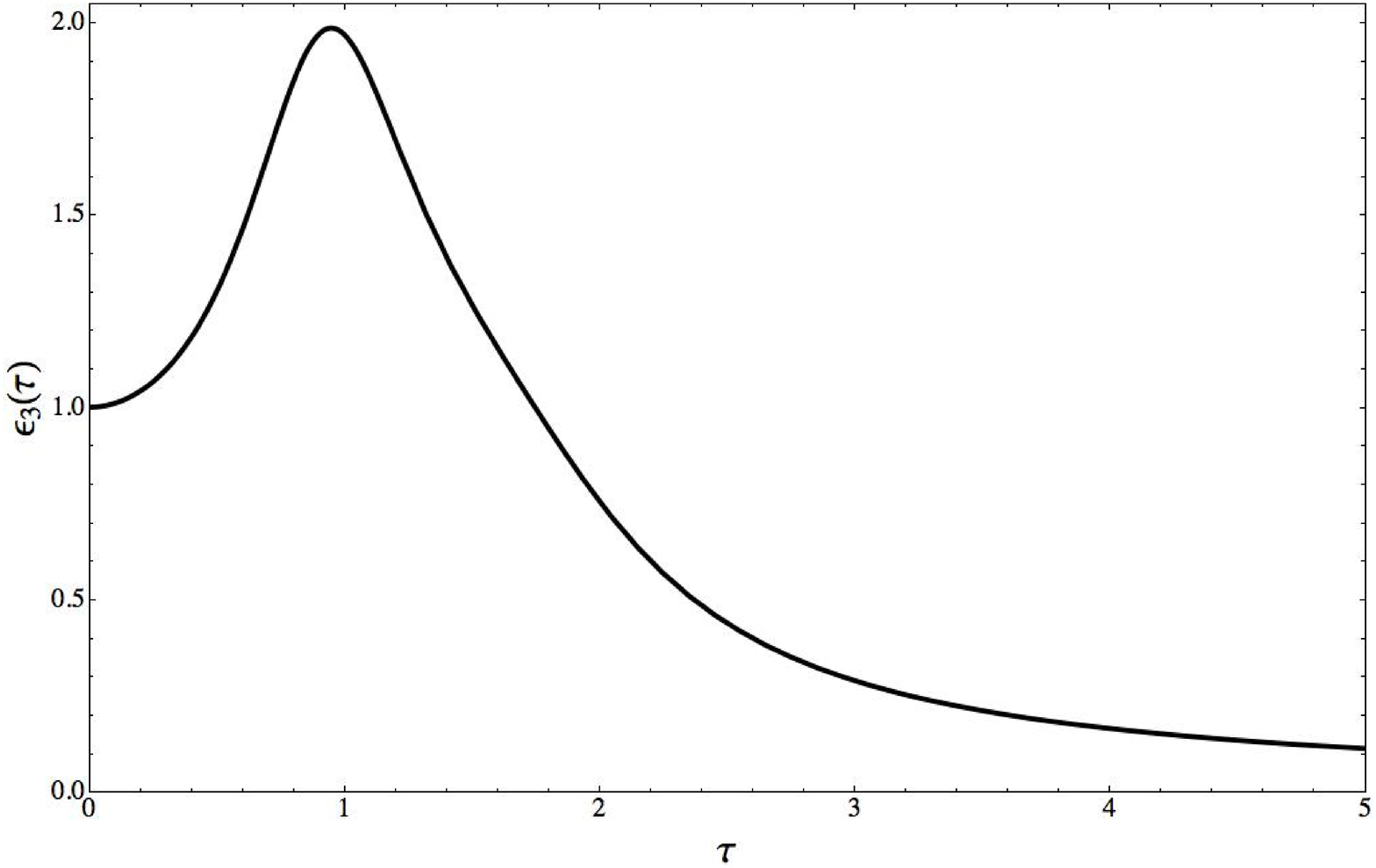}
\caption{Pad\'e resumed profiles for $\epsilon(\tau)$ for the initial conditions $v +w = \tanh(z^2) - \tan (z^2)$, $v+w = \tanh(z^2 + z^8 /6) - \tan(z^2)$ and 
$v+w= 2/3 z^6 (1+z^2/2)/(z^2-1)$, respectively.}
\label{fig:epsilon}
\end{figure}

One can also perform the Pad\'e approximation assuming the late time exponent $s=4/3$. The profiles of $\epsilon(\tau)$ obtained in this way are plotted in Fig.\ref{fig:epsilon} 
for a set of initial conditions (plots taken from \cite{Beuf:2009cx}). The energy densities differ in the initial stages of the evolution, whereas in the late-time regime they seem to approach local 
equilibrium.

Recall that the positivity condition $T_{\mu \nu} t^\mu t^\nu \ge 0$ implied $- 4 \epsilon / \tau \le \epsilon^{\prime} \le 0$ (see eq. (\ref{epsiloncond})).  
In the third plot of Fig.\ref{fig:epsilon}, a temporary violation of the positive energy condition is observed. 
Such a transient behavior may appear for a quantum-driven process (see $e.g.$ the discussion in \cite{grumi}, quoting field-theoretical results \cite{field}.).

\begin{figure}[t]
\centering
\includegraphics[width=0.45\linewidth]{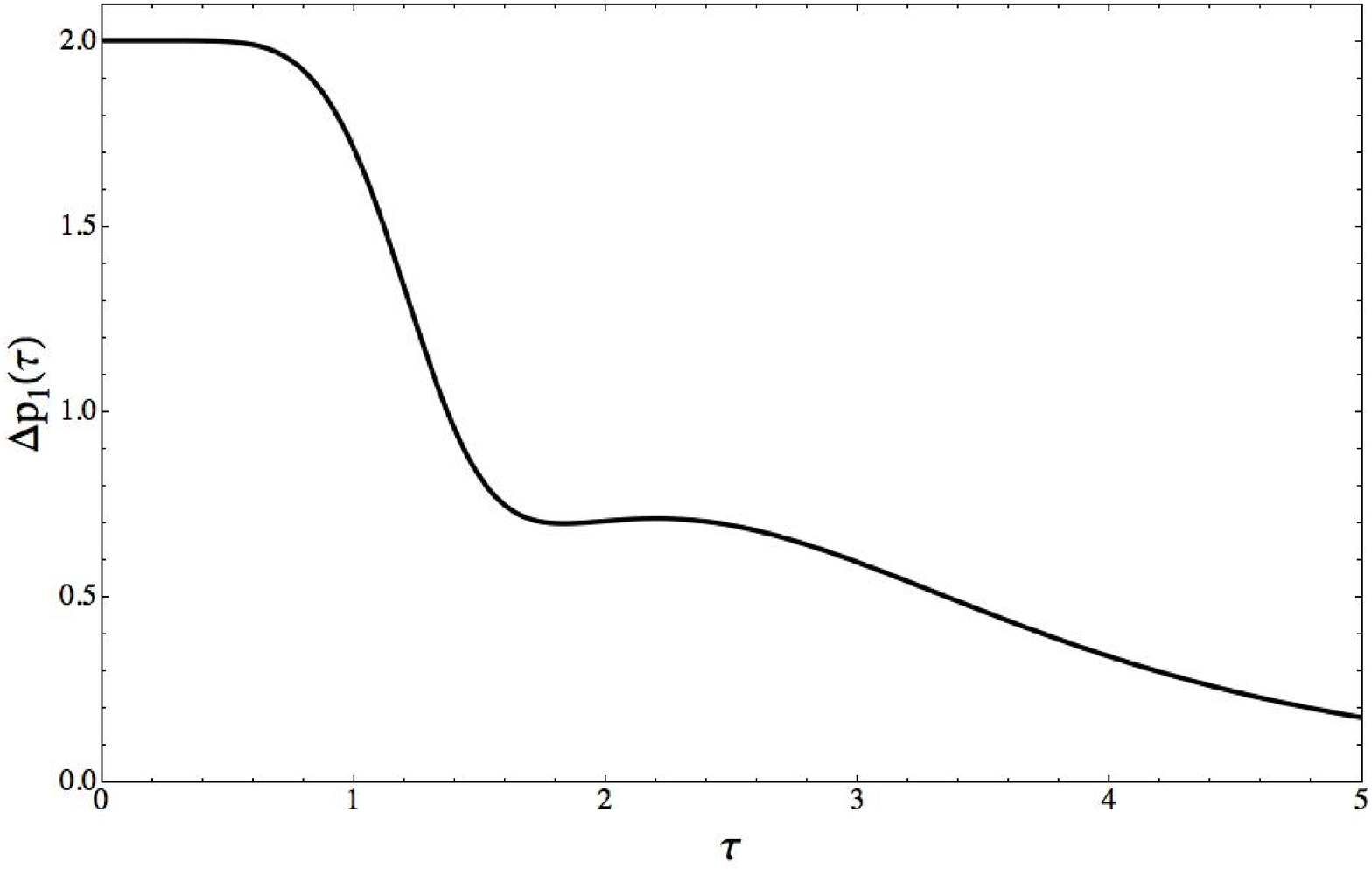}
\includegraphics[width=0.45\linewidth]{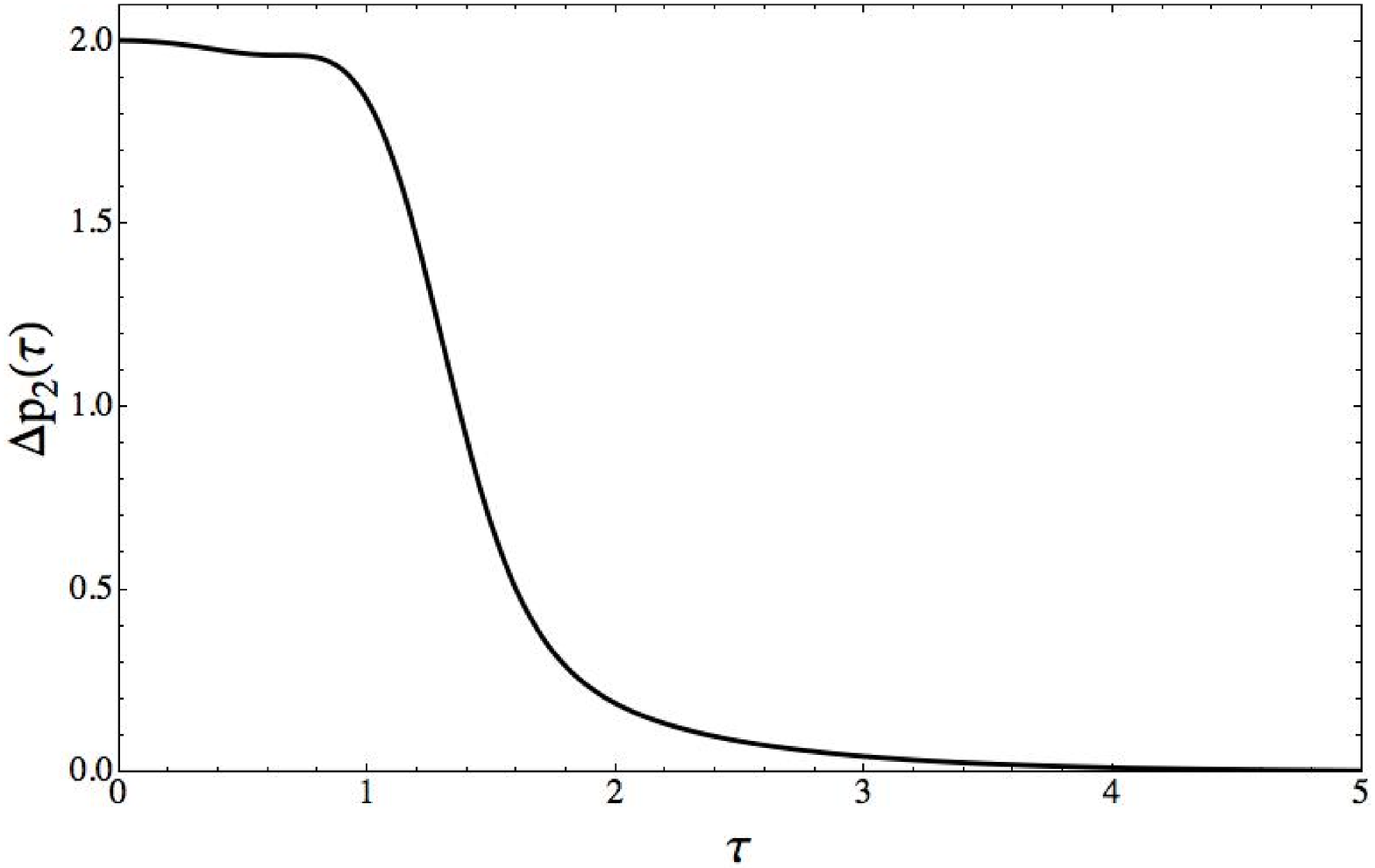}
\caption{Relative difference in pressures $\Delta p(\tau)= 1 - p_{\parallel}/p_{\perp}$ for the initial conditions $v +w = \tanh(z^2) - \tan (z^2)$ and $v+w = \tanh(z^2 + z^8 /6) - \tan(z^2)$, respectively.
The profiles exhibit a rapid fall-off at $\tau \sim \mathcal O (1)$, but the first profile does not reach yet complete isotropization. }
\label{fig:deltap}
\end{figure}

An alternative way to describe the transition to hydrodynamics is through the relative difference between the longitudinal and transverse 
pressures defined as
\be\label{deltap}
\Delta p(\tau) = 1 - \frac{p_{\parallel}(\tau)}{p_{\perp} (\tau)}\,.
\ee
When the quantity $\Delta p(\tau)$ approaches 
zero, it signals isotropization indicating local equilibrium, while a value of order one indicates a situation close to a free streaming scenario. Fig.\ref{fig:deltap} 
appears in \cite{Beuf:2009cx} and
shows the relative difference of pressures (\ref{deltap}) for a couple of initial conditions. It is interesting to observe that both profiles exhibit a rapid fall-off on a 
scale $\tau \sim \mathcal O (1)$. In the first profile however, isotropization remains incomplete until $\tau \sim \mathcal O (5)$.  It is interesting to note that figures \ref{fig:epsilon} 
and \ref{fig:deltap} are  consistent with the full anisotropy solution at {\it very small} proper-times but show a variety of behavior in the whole of the early proper-time region. 

These results point towards a fast isotropization, which could however remain incomplete for some time. 
The convergence to the hydrodynamic regime remains an open problem. In the coming more systematic numerical study \cite{Progress} the stability of the Pad\'e approximation has been shown reliable to a larger proper-time interval through non trivial reparametrization of the metric. A more general proof of the transition to the hydrodynamic regime, filling the gap between early-time and late-time behavior is still lacking. Some progress has been made, as we shall see now, starting from specific {\it shock wave} initial conditions mimicking the initial ultra-relativistic colliding heavy-ions.


\subsection{Shock wave initial conditions}\label{ShockWaves}

An interesting way to model ultra-relativistically boosted large nuclei is through gauge theory shock waves. In this context, by shock wave we mean a plane shell of matter moving at 
the speed of light. The gravity dual of a boundary shock wave has been first constructed in \cite{Janik:2005zt} and later applied to various analysis, 
such as in attempts to describe holographically deep inelastic scattering or heavy-ion reactions through shock waves collisions. 

In terms of the light-cone coordinates $x^{\pm} \equiv t \pm y$, an ultra relativistic source extended in the transverse direction (a model for an incident heavy ion with large transverse size) and moving along the direction $x^+$   has an energy-momentum tensor
\be\label{deltaT}
T_{--} \equiv F(x^-) \sim \mu \, \delta(x^-)\,.
\ee
The last term represents more specifically a shock wave, $i.e.$ a shell with vanishing thickness and
 transverse energy density $\mu$. It may be more concretely represented by a Gaussian centered at $x^-=0.$

For arbitrary profile $F(x^-),$ the dual background to this configuration reads 
\be\label{ShockWave}
z^2 ds^2 \!=\! - dx^- dx^+\!\! +  z^4 F(x^-) dx^{-2}\! + d\vec x_{\perp}^2\! + dz^2\!,\ 
\ee
where $F(x^-) = \langle T_{--} \rangle$, and is an exact solution\footnote{Note that the  solution (\ref{ShockWave}) has been provided a generalization to profiles $F = F(x^-, \vec x_{\perp}, z)$ in \cite{Beuf:2009mk}.} to Einstein equations \cite{Janik:2005zt}.

\begin{figure}[t]
\begin{center}
\includegraphics[scale=0.25]{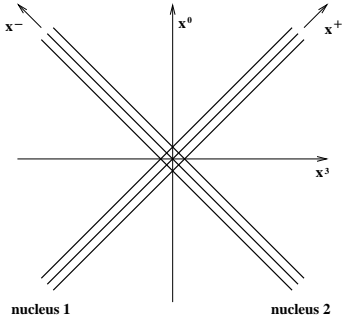}
\caption{Space-time diagram of the collision of two ultra-relativistic nuclei in the center of mass frame.}
\label{fig:shockwaves}
\end{center}
\end{figure}

These solutions constitute the starting point towards solving the problem of shock waves collisions. Shock waves having a very small thickness along the 
collision axis and localized in the transverse directions qualitatively resemble the highly Lorentz-contracted nuclei in a heavy-ion collision. 
Fig.\ref{fig:shockwaves} appeared 
in \cite{Albacete:2008vs} and depicts the space-time picture of ultra-relativistic heavy-ion collisions in the center of mass frame.  
In the AdS/CFT language, the collision of shock waves in $\mathcal N=4$ SYM translates into colliding gravitational shock waves in the bulk. 

Several deserving attempts, both analytical and numerical, have been made to solve this problem 
\cite{grumi,Albacete:2008vs,Albacete:2009ji,Kang:2004jd,Gubser:2007zr,Gubser:2008pc,AlvarezGaume:2008fx,Lin:2009pn,Gubser:2009sx,Kovchegov:2009du}, 
but the full understanding of the transition to the hydrodynamic regime is still an active subject of research. Let us discuss some of these angles of attack.

The collision of two singular shock waves moving in the directions $x^{\pm}$ on the boundary of a five-dimensional AdS space was studied in \cite{grumi}. 
The pre-collision line element is given by the superposition of two metrics of the form (\ref{ShockWave}), with delta-function Ansatz $F(x^{\pm}) \sim \delta(x^{\pm})$. Solving Einstein's equations 
in a power series expansion in the proper-time allows to extract the expression of the energy-momentum tensor right after the collision. In \cite{grumi}, 
it was observed that the latter quantity is not boost-invariant in the sense of Bjorken. Boost-invariance could be eventually re-established at later times, 
at which the expansion breaks down. A crude estimate of the thermalization time in this simple model leads to
$\tau_{therm} \sim \mu^{- \frac 1 3}$. 
The non-trivial dimensionless factor in the above relation is $\mathcal O (1)$. 

In \cite{Albacete:2008vs}, it was observed that the metric of a single shock wave, moving along the light cone, can be computed exactly in first order perturbation 
theory. The only diagram contributing represents a single graviton exchange between the source nucleus at the AdS boundary and the location in 
the bulk where the metric is measured. The metric after the collision in the forward light cone can then be constructed perturbatively through an expansion 
in graviton exchanges. This computation was carried out at tree level in \cite{Albacete:2008vs} and up to NNLO in \cite{Albacete:2009ji}, where the 
technique was applied to proton-nucleus collisions. 

The analysis of the holographic stress-energy tensor leads to the strong prediction of full nuclear stopping of the two nuclei shortly 
after the collision. This happens after a time of the order of the inverse typical transverse momentum scale, independently of the energy density. 
This behavior was interpreted in \cite{Albacete:2008vs} as signaling black hole creation in the bulk. If the two nuclei stop completely, the strong interactions exerted 
between them are likely to thermalize the system, eventually leading to Landau hydrodynamics at late times \cite {lan}. This non boost-invariant solution is not expected to satisfy the phenomenological constraints, even if the description of multiplicities seems to be valid. 

In  \cite{Albacete:2008vs,Albacete:2009ji} it was
pointed out that weak interactions are expected to be determinant in the very early stages of the collision \cite {CGC?}. Since a realistic model should take them into account, and in absence of a dual description incorporating weak coupling aspects of QCD,
it was proposed to mimic these effects through net zero-energy unphysical shock waves with profiles $F(x^{\pm}) \sim \delta^{\prime}(x^{\pm})$. The resulting 
energy density of the strongly coupled medium starts out as a constant at early proper time, in agreement with the results of \cite{Kovchegov:2007pq}.      

The above analytic approaches have the limit of not being able to address the non-linearity of the thermalization process. A numerical 
treatment of the full problem appeared in \cite{Chesler:2010bi}, after these lectures were given.\footnote{Other recent papers related to our lectures are quoted in Refs.\cite{ingo,iizuka,alice}.} 
The main advantage of solving the initial 
value problem numerically is that it is more likely to properly describe the transition between the early and late time behaviors.

Ref. \cite{Chesler:2010bi} considered the collision of two planar sheets in $\mathcal N=4$ SYM propagating towards each other at the speed 
of light, with finite energy density, finite thickness and Gaussian profile
\be\label{FCY}
F(x^{\pm}) = \frac{\mu^3} {\sqrt{2 \pi \omega^2}} e^{- \left({x^{\pm}}/{\om \sqrt 2}\right)^2}\,.
\ee
The equivalent gravitational problem is the collision of two planar shock waves which are regular, non-singular, source-less solutions 
to Einstein's equations. The collision leads to horizon formation in the bulk and the numerical integration was stopped in \cite{Chesler:2010bi} at the apparent horizon location. 

The energy density of the collision products is found to be peaked around two receding maxima which move outwards at less than the speed of light. 
Setting the width in (\ref{FCY}) to $\omega = 0.75/\mu$, the application of this model to the RHIC collisions leads to $\mu \sim 2.3$ GeV.  The total 
time required for thermalization, from when the Gaussian shock waves start to overlap significantly until the onset of hydrodynamics, is estimated to be 
$\Delta v_{tot} \sim 4/\mu \sim 0.35$ fm/c ($v$ denotes time in ingoing Eddington-Finkelstein coordinates). This timescale is consistent 
with the results of \cite{Chesler:2009cy}.



\section*{Acknowledgements}
A.B. would like to thank the lecturers, organizers and participants of the Carg\`ese Summer School for providing an active environment for lively and valuable discussions.  
The research of A.B. is supported by the Belgian Federal Science Policy Office through the Interuniversity Attraction Pole IAP VI/11 and by FWO-Vlaanderen through project G011410N. 
R.P. would like to mention the continuously inspiring collaboration with Romuald Janik on the research subject of these lectures. He also thanks the organizers for the invitation and the participants for the stimulating and friendly atmosphere of the Carg\`ese summer school. We thank Rudolf Baier for reminding the relevant Ref.\cite{baier}. 

\end{document}